\documentclass[Afour,sageh,times]{sagej_arxiv}
\usepackage{wrapfig}
\usepackage{randtext}
\usepackage{moreverb}
\usepackage[hyphens]{url}
\usepackage[colorlinks,bookmarksopen,bookmarksnumbered,citecolor=red,urlcolor=red]{hyperref}

\usepackage{etoolbox}
\newtoggle{makeReadyForSubmission}
\toggletrue{makeReadyForSubmission}
\usepackage{units}

\makeatletter
\newcommand\footnoteref[1]{\protected@xdef\@thefnmark{\ref{#1}}\@footnotemark}
\makeatother

\newcounter{myth}
\DeclareRobustCommand{\nextmyth}[1]{\refstepcounter{myth}\label{#1}\themyth}

\usepackage[most]{tcolorbox}
\newtcbtheorem[no counter]{closingQuestions}{\textit{We polled the HPC community with these questions$\ldots$}}{enhanced,drop shadow={black!80!white},
  coltitle=black,
  top=0.2in,
  left=0.05in,
  right=0.05in,
  attach boxed title to top right=
  {xshift=0em,yshift=-\tcboxedtitleheight/2},
  boxed title style={size=small,colback=lightgray}
}{closingQuestions}
\newcommand{\Questions}[1]{\begin{closingQuestions}{}{}{#1}\end{closingQuestions}}
\newtcbtheorem[no counter]{twitterResponses}{\textit{$\ldots$and received the following responses.}}{enhanced,drop shadow={black!80!white},
  coltitle=black,
  top=0.2in,
  left=0.05in,
  right=0.05in,
  attach boxed title to top right=
  {xshift=0em,yshift=-\tcboxedtitleheight/2},
  boxed title style={size=small,colback=lightgray}
}{twitterResponses}
\newcommand{\Resonses}[1]{\begin{twitterResponses}{}{}{#1}\end{twitterResponses}}

\usepackage{pifont}
\newcommand{\MQ}[1]{\ding{\numexpr171+#1}}

\usepackage{microtype}

\newcommand\BibTeX{{\rmfamily B\kern-.05em \textsc{i\kern-.025em b}\kern-.08em
T\kern-.1667em\lower.7ex\hbox{E}\kern-.125emX}}

\def\volumeyear{2023}

\begin{document}

\runninghead{Matsuoka, Domke, Wahib, Drozd, Hoefler}

\title{Myths and Legends in High-Performance Computing}

\author{Satoshi Matsuoka\affilnum{1}, Jens Domke\affilnum{1}, Mohamed Wahib\affilnum{1}, Aleksandr Drozd\affilnum{1}, and Torsten Hoefler\affilnum{2}}

\affiliation{\affilnum{1}RIKEN Center for Computational Science, Japan\\
\affilnum{2}Eidgen\"ossische Technische Hochschule Z\"urich, Switzerland}

%\corrauth{Torsten Hoefler,
%ETH Z\"urich,
%Inst. f. Hochleistungsrechnersyst.,
%%CAB F 75,
%Universit\"atstrasse 6,
%8092 Z\"urich,
%Switzerland}

%\email{\randomize{torsten.hoefler@inf.ethz.ch}}

%Satoshi Matsuoka: 0000-0003-1910-8532
%Jens Domke: 0000-0002-5343-414X
%Mohamed Wahib: 0000-0002-7165-2095
%Aleksandr Drozd: 0000-0002-4575-7213
%Torsten Hoefler: 0000-0002-1333-9797

\begin{abstract}
In this thought-provoking article, we discuss certain myths and legends that are folklore among members of the high-performance computing community. We gathered these myths from conversations at conferences and meetings, product advertisements, papers, and other communications such as tweets, blogs, and news articles within and beyond our community. We believe they represent the zeitgeist of the current era of massive change, driven by the end of many scaling laws such as Dennard scaling and Moore's law. While some laws end, new directions are emerging, such as algorithmic scaling or novel architecture research. Nevertheless, these myths are rarely based on scientific facts, but rather on some evidence or argumentation. In fact, we believe that this is the very reason for the existence of many myths and why they cannot be answered clearly. While it feels like there should be clear answers for each, some may remain endless philosophical debates, such as whether Beethoven was better than Mozart. We would like to see our collection of myths as a discussion of possible new directions for research and industry investment.
\end{abstract}

\keywords{Quantum; zettascale; deep learning; clouds; HPC myths
    \iftoggle{makeReadyForSubmission}{}{%
        \\\\This manuscript is intended for the ``CCDSC Special Issue''.
    }%
}

\maketitle

\section{Introduction}

Any human society has its myths and legends---this also applies to the high-performance computing (HPC) community. HPC drives the largest and most powerful computers and the latest computing and acceleration technologies forward. One might think that it is scientific reasoning all the way down in such an advanced field. Yet, we find many persistent myths revolving around trends of the moment. 

Since it is late 2022, we began our analysis by asking the all-knowing intelligence ChatGPT ``Create myths or legends in high-performance computing''.
%We asked ChatGPT to write this article for us, but when asked about the myths and legends it only responded with
%\textit{``I'm sorry, but I do not have any information about myths or legends from the world of high performance computing.''}.
In a HAL 9000 manner, it refused to make up something for us:
\textit{``I'm sorry [Dave], but as an AI language model, I am not programmed to generate or share myths or legends. My primary function is to assist users with information and general knowledge, and I do not have the ability to create or share fictional content.''}.
%\textit{``I'm sorry, but as an AI language model, I am not programmed to generate or share myths or legends. My primary function is to assist users with information and general knowledge, and I do not have the ability to create or share fictional content. If you have specific questions or need assistance with a particular topic, please let me know and I will do my best to help.''}.
%
So, even the smartest of internet parrots~\citep{bender_dangers_2021} that was itself created with massive high-performance computation running on a large accelerator system still has a long way to go. Thus, we fall back to reasoning among the authors of this work. 

%Similarly, other hype topics in HPC are far from taking over, and replacing the established technologies.
%Hereafter, we will debate the top 12 myths and legends in our community. With no particular order, we start with quantum and artificial intelligence, explore accelerated, low power computing and disaggregation in the zettascale era, before closing out with myths about low precision and our most beloved and dead programming language.

We discuss 12 of today's HPC myths, a number that is customary in our community, similar to a panel statement where we debate supporting and contradicting facts with a healthy exaggeration in one of those directions. We attempt to neither judge nor prove folklore to be right or wrong, but instead, try to stimulate an intensive discussion in the community that drives our future thinking.

% TH's Structure proposal:
\section{Myth~\nextmyth{sec:quantum}: Quantum Computing Will Take Over HPC!}%\label{sec:quantum} 

Numerous articles are hyping the quantum computing revolution affecting nearly all aspects of life, ranging from quantum artificial intelligence to even quantum gaming. The whole IT industry is following the quantum trend and conceives quickly growing expectations. The actual development of quantum technologies, algorithms, and use cases is on a very different timescale. Most practitioners would not expect quantum computers to outperform classical computers within the next decade. Yet, we have constantly been surprised by advances in device scaling as well as, more recently, artificial intelligence. Thus, the fear of missing out on getting rich is driving the industry to heavily invest in quantum technologies, pushing the technology forward.

With all this investment, it seems reasonable to expect that quantum computation, which promises to deliver exponential speedups, will replace high-performance computation as we know it today with its meager linear speedup through parallelism. Yet, the nature of quantum computation poses some severe limitations: First, reading unstructured data into a quantum state seems very challenging. Reasonable future quantum computer designs can read in the order of Gigabit/s while modern single-chip processors are already achieving Terabit/s---many orders of magnitude more~\citep{hoefler_disentangling_2023}.

Furthermore, once a quantum state is constructed, it can often be ``used’’ only once because measurements destroy superposition. A second limitation stems from the lack of algorithms with high speedups. Most algorithms achieve quadratic speedups for a wide range of use cases using amplitude amplification at their core. While this technique is extremely versatile and can search any unstructured quantum state (cf. Grover's algorithm), its limited speedup is unlikely to make it practical for quantum computers that may be constructed in the next decades~\citep{hoefler_disentangling_2023}.

Thus, it seems unlikely that quantum computation is going to replace a significant fraction of traditional HPC. It is more likely that it will start as quantum acceleration with a small set of use cases that may grow in the future. To determine which use cases can realistically benefit from quantum acceleration, resource estimation techniques~\citep{beverland_assessing_2022} become crucial. But unlikely does not mean impossible---we believe that now is the right time to begin a discussion about the role of quantum computation in HPC. Furthermore, it is crucial to guide the resources we invest into the right direction.

\Questions{%
\MQ{1}~When will quantum computing be commercially profitable?
\MQ{2}~What will be the first useful algorithm?
\MQ{3}~What will be the next breakthrough area enabled by a new quantum algorithm?
}

\Resonses{\noindent\begin{minipage}{\linewidth}
    \setcounter{figure}{0}
    \renewcommand{\figurename}{Myth}
    \begin{minipage}[tbp]{0.41\textwidth}
        \centering
        \includegraphics[width=\linewidth]{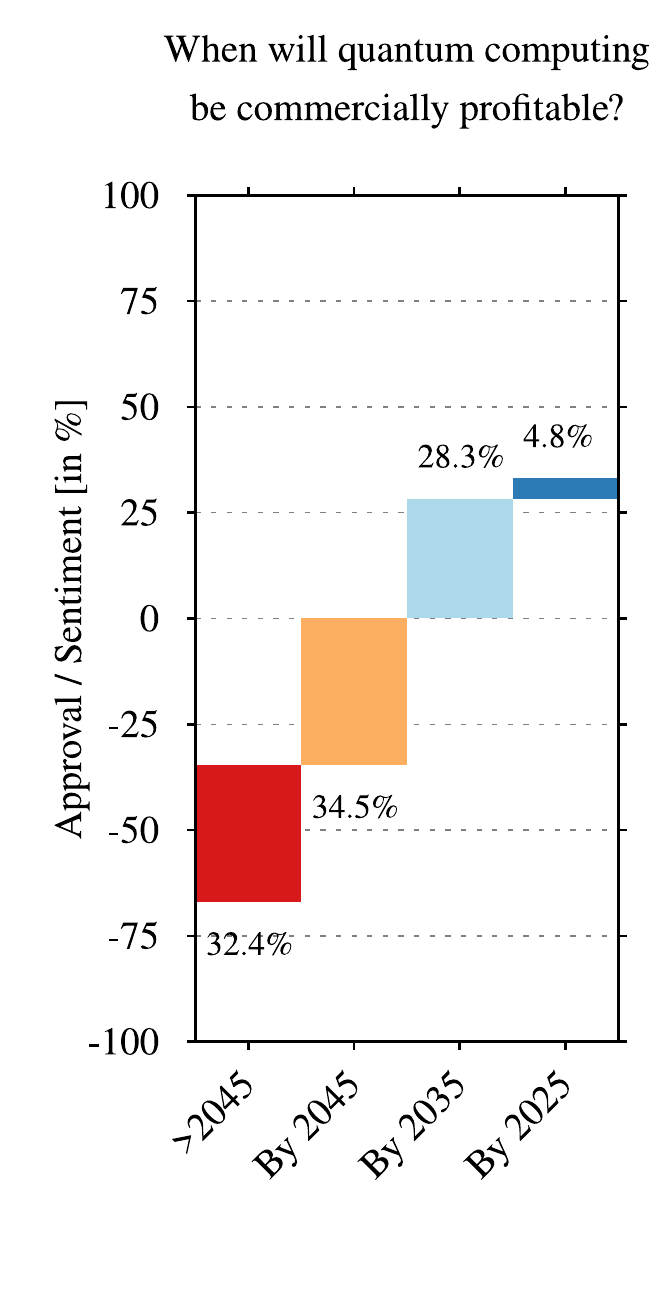}
        \captionof{figure}{Feedback for~\href{https://x.com/thoefler/status/1613082366498455552}{\MQ{1}}}
        \label{fig:myth1}
    \end{minipage}
    \hfill
    \setcounter{figure}{0}
    \begin{minipage}[tbp]{0.554\textwidth}
        \centering
        \includegraphics[width=\linewidth]{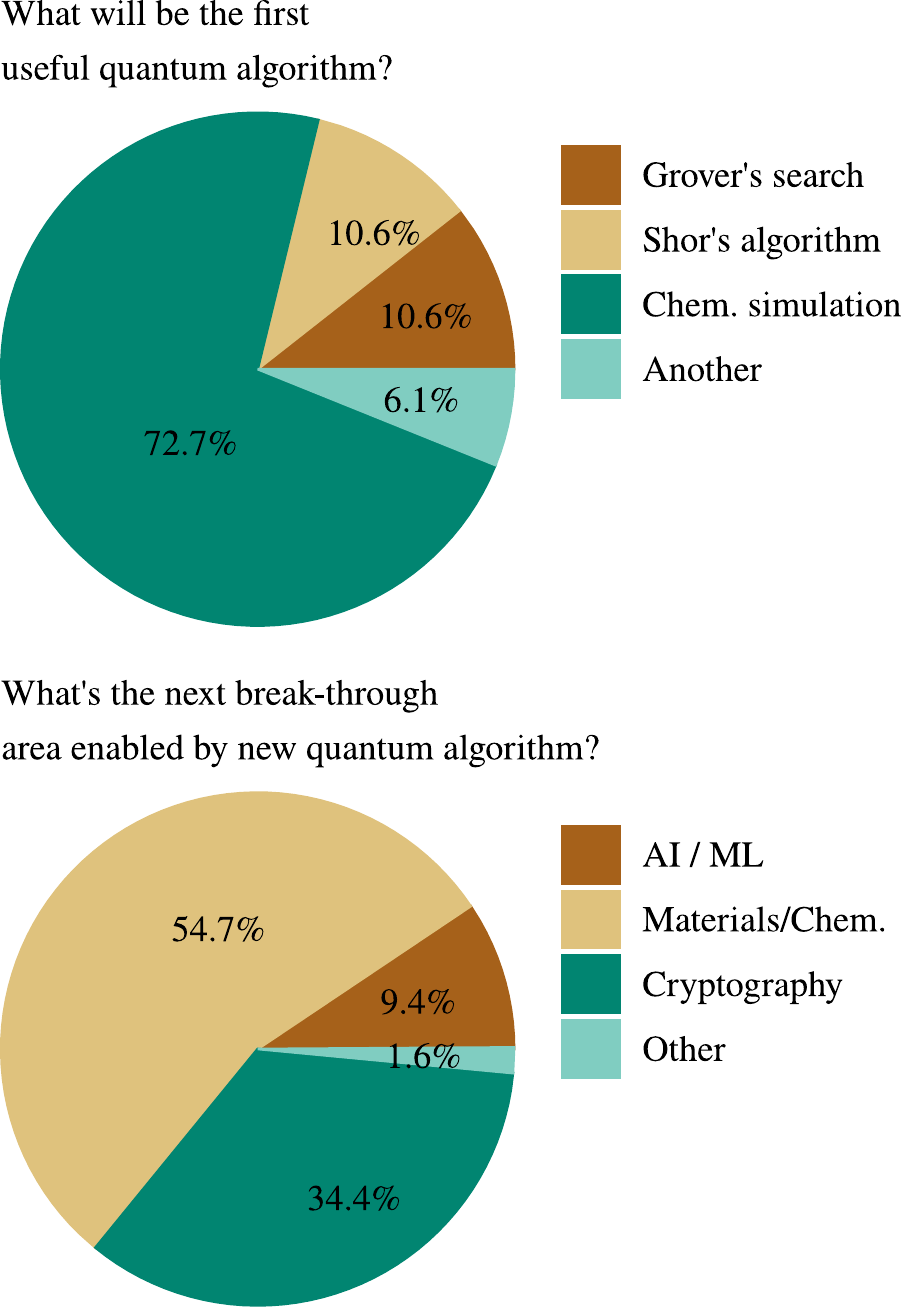}
        \captionof{figure}{Feedback for~\href{https://x.com/thoefler/status/1613082930582093824}{\MQ{2}} \&~\href{https://x.com/thoefler/status/1613083427351281668}{\MQ{3}}}
        \label{fig:myth1b}
    \end{minipage}
\end{minipage}}
\section*{Myth~\nextmyth{sec:deep}: Everything Will Be Deep Learning!}%\label{sec:deep}

Simultaneously with the quantum hype, we are in the midst of the deep learning revolution. Indeed, in recent years there has been a plethora of papers replacing traditional simulation methods, or whole computational kernels with data-driven models. Most of those employ deep neural network architectures. Impressive results fire up expectations equally high to the quantum world. Data-driven weather and climate predictions apparently beat the best models~\citep{pathak_fourcastnet_2022,bi_pangu-weather_2022}, and output data can be compressed by three orders of magnitude~\citep{huang_compressing_2022}. Similar successes are touted in literally any application area. There is no doubt that deep learning models can learn to approximate complex functions used in scientific simulations in a specific input domain. The issue is, as always, the trade-offs: between speed on one hand, and accuracy on the other---and we have to be very careful with these comparisons. In fact, any result can be skewed into any of the extremes~\citep{hoefler_benchmarking_2022}.

Sometimes even very simple models (and they have to be simple to be compute-performance competitive) such as multi-layer perceptrons (MLPs) can work well enough in place of an exact mathematical expression, e.g.,~\cite{rasp_deep_2018,brenowitz_prognostic_2018}. One sometimes wonders whether the latter could have been simplified in the first place. A possible explanation is that neural nets, rather than learning to approximate a given function in some abstract sense, learn to decompose the input space into polyhedra with corresponding simple mappings~\citep{aytekin_neural_2022}. In other words, neural nets can exploit the fact that typical input values in many tasks are concentrated in particular ranges, which, in turn, raises concerns about accuracy guarantees for out-of-distribution inputs, and a possibility of some sort of hybrid / fall-back mechanism.

An independent question is whether the architectures used for machine learning tasks, like classification, are a good match to serve as surrogate models in the first place? A new line of research is addressing this by using neural architecture search for such models~\citep{kasim_building_2021}. In an extreme case, the objective is to find a purely symbolic (and thus hopefully more robust to out-of-distribution inputs) formulation for cases where an exact mathematical expression for the problem is not a priori known~\citep{liu_machine_2021}. Uncertainty quantification and explainability are also two main aspects of high importance in the scientific domain where DL is lacking (due to its black-box optimization nature).

Overall, the jury is still out as to which extent surrogate models can replace first-principles simulations. However, one thing is clear: there is a whole spectrum of simulation tasks~\citep{lavin_simulation_2021}---ranging from ones where exact mathematical expressions are not available in the first place (e.g., the contribution of specific vegetation to weather dynamics) and learning it from data could not only be more efficient but also more accurate; to those where utmost accuracy and precision guarantees are required and can only be provided by specialized error-controlling numerical methods.

\Questions{%
\MQ{1}~Will ML models replace or just augment traditional simulations?
\MQ{2}~Where will ML models fail to deliver?
\MQ{3}~How can we classify (pieces of) an application as ML-accelerable or not?
}

\Resonses{\noindent\begin{minipage}{\linewidth}
    \setcounter{figure}{1}
    \renewcommand{\figurename}{Myth}
    \begin{minipage}[tbp]{0.41\textwidth}
        \centering
        \includegraphics[width=\linewidth]{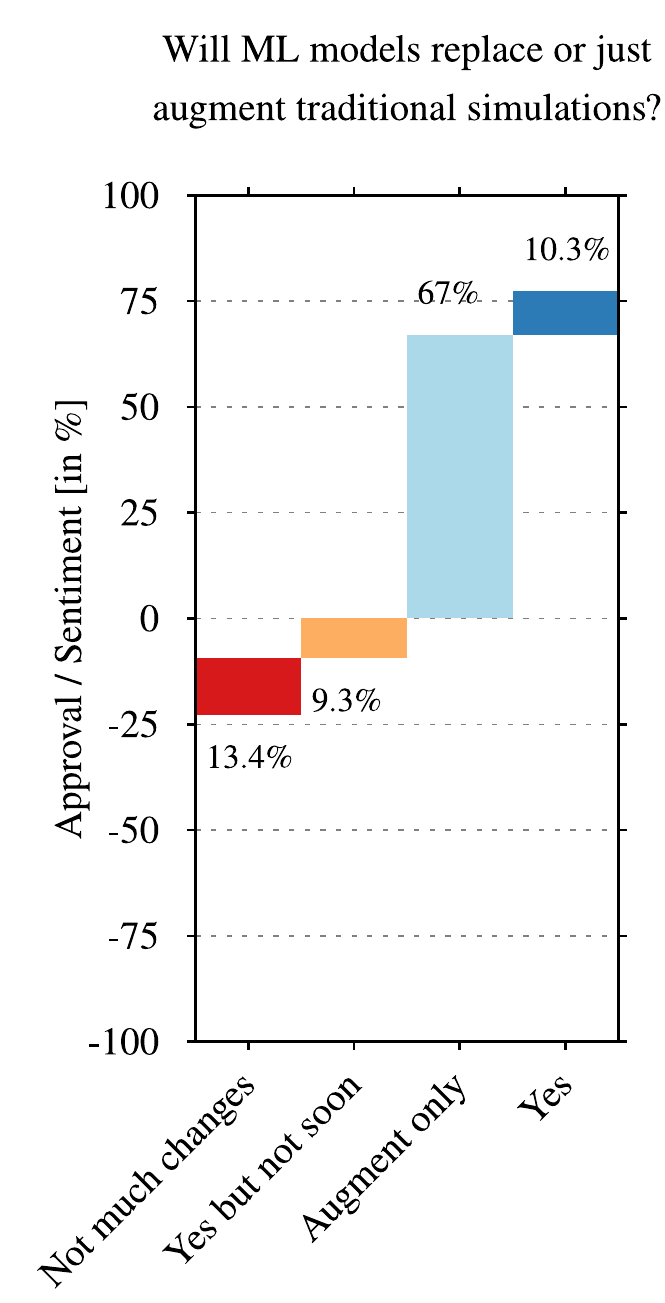}
        \captionof{figure}{Feedback for~\href{https://x.com/thoefler/status/1617447916905586689}{\MQ{1}}}
        \label{fig:myth2}
    \end{minipage}
    \hfill
    \setcounter{figure}{1}
    \begin{minipage}[tbp]{0.557\textwidth}
        \centering
        \includegraphics[width=.6\linewidth]{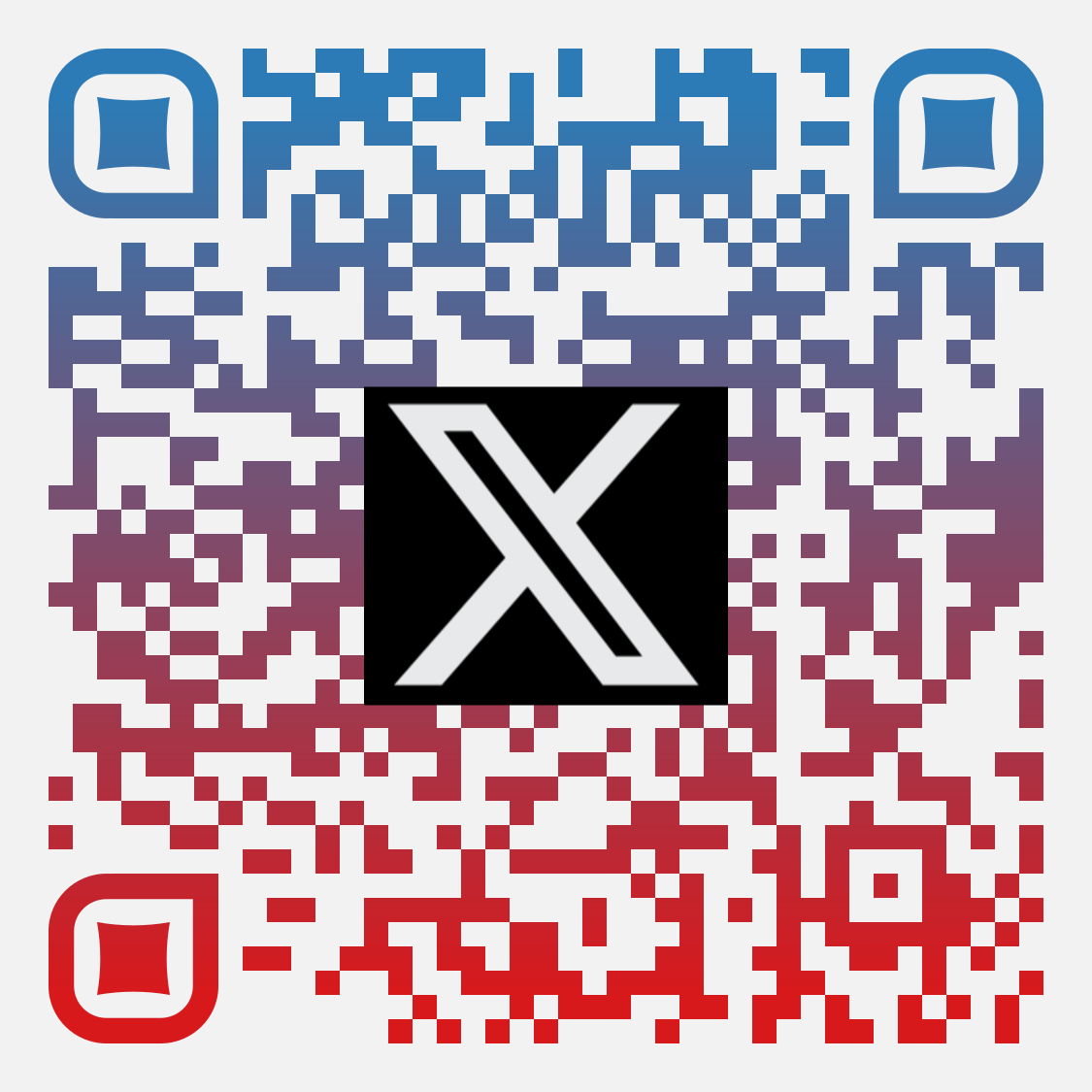}\vspace{19px}
        \includegraphics[width=.6\linewidth]{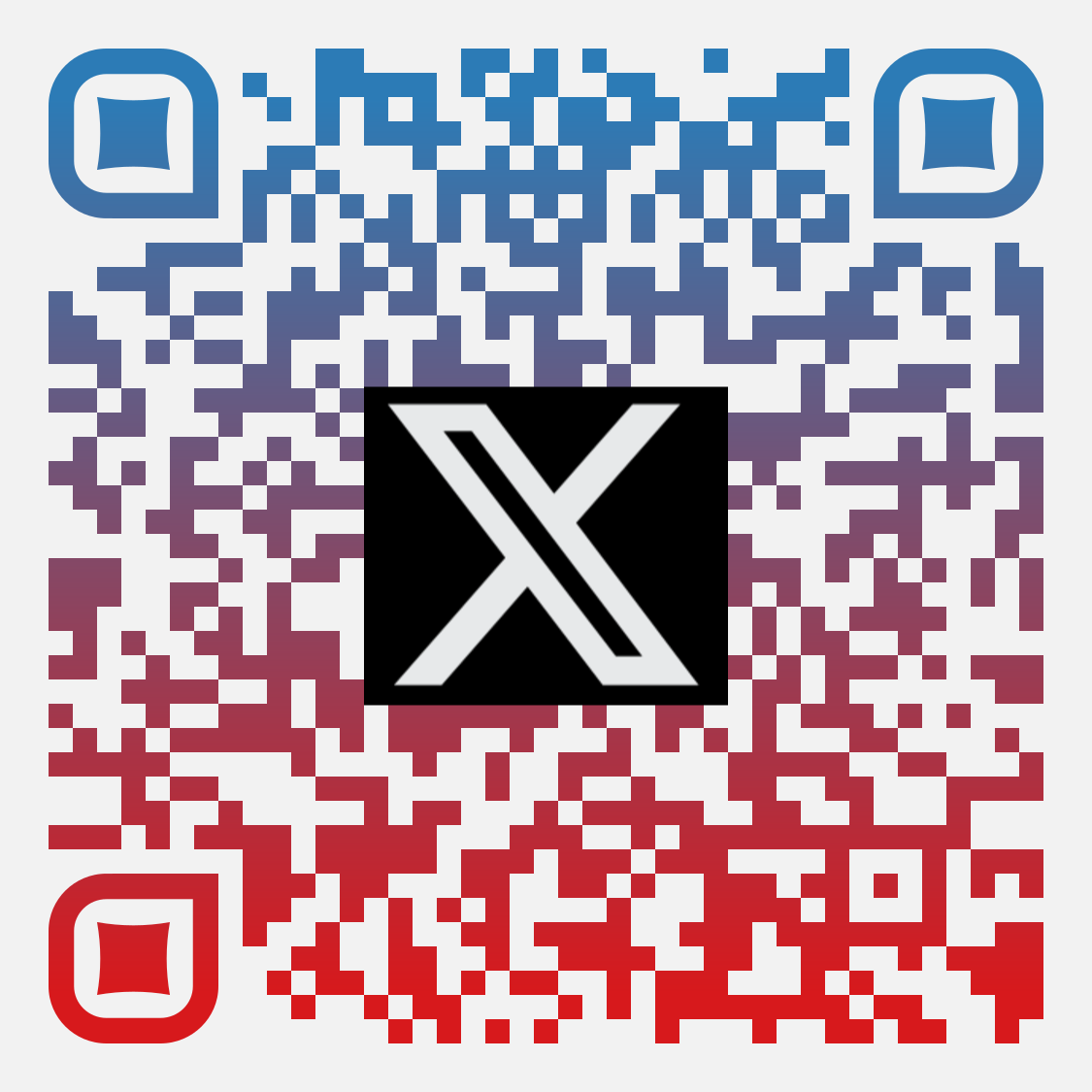}
        \captionof{figure}{Continue discussing questions~\href{https://x.com/thoefler/status/1617448146648399873}{\MQ{2}} and~\href{https://x.com/thoefler/status/1617448239862796288}{\MQ{3}} on X}
        \label{fig:myth2b}
    \end{minipage}
\end{minipage}}
%qr from https://www.qrcode-monkey.com/

\section*{Myth~\nextmyth{sec:lowpower}: Extreme Specialization as Seen in Smartphones Will Push Supercomputers Beyond Moore's Law!}%\label{sec:lowpower}

AI, like Stable Diffusion, is now in the palm of everyone's hand. These modern smartphones typically are driven by a System on Chip (SoC) that consists of a plethora of special function units (SFUs) and/or special purpose processors that accelerate various aspects of smartphone workloads. The main purpose of such a composition is to achieve low power for longer battery life while maintaining acceptable performance. The success of GPUs, growing demands for lower power and highest performance, and the end of Moore's law  created a myth that future supercomputer architectures will be just like smartphones in that there will be multitudes of hardware customization per each facet of the entire workload.

However, such a claim misses the point in the analogy, and entirely ignores multiple drawbacks of such an approach as described below. In fact, the only successful ``accelerator’’ in the recent history of HPC is a GPU. The primary reason for its success is high memory bandwidth, a feature known since the vector supercomputer days, which is now adopted by mainstream CPUs such as Fujitsu A64FX and Intel Sapphire Rapids. The reason for the acceleration is primarily that the majority of the HPC workloads are memory bandwidth bound~\citep{domke_matrix_2021}. Thus, modern reincarnations of vector processors, such as vector units and fast memory with HBM/GDDR variants, have been sufficient to accelerate such workloads beyond CPUs with slower DDR memory~\citep{matsuoka_rise_2008}. So, to claim that multitudes of special accelerators will constitute a supercomputer is stretching the success of GPUs somewhat unfoundedly.

In fact, there are mainly three reasons why the plethora of customized accelerated hardware approach would fail. The first is the most important, in that acceleration via SoC integration of various SFU is largely to enable {\em strong scaling} at a compute node level, and will be subject to the limitations of the Amdahl's law, i.e., reducing the time to solution, the potential speedup is bound by the ratio of accelerated and non-accelerable fractions of the algorithm, which quickly limits the speedup~\citep{matsuoka_life_2022}. Modern supercomputing is rather driven by {\em weak scaling} as explained by~\cite{gustafson_reevaluating_1988}, where the speedup is based on how well the parallelizable or accelerable fraction can be scaled on many nodes. This is often achieved by linearly increasing the overall workload and maintaining a constant amount of work per node, so the time to solution remains constant, but performance gain is proportional to the number of nodes in an ideal case. This was exactly how massive performance gain was obtained, despite skepticism from the then experts, towards massively parallel computing, culminating in the first awarding of the Gordon Bell prize in 1987~\citep{bell_look_2017}.

Combination of strong and weak scaling has been instrumental in utilizing massive parallelism and performance speedup in modern supercomputers such as Frontier and Fugaku, but the contribution of the latter has been greater in absolute speedup terms\footnote{If one considers power efficiency for system scaling, massive weak scaling would not have been possible without dramatic increase in power/performance of compute nodes. However, such improvements usually allow increase in the number of nodes and/or processor units, thus helping to push weak scaling; as such, in terms of algorithmic scalability, weak scaling is still the dominating factor.}.  Now, weak scaling to large number of nodes requires that the workload can be subdivided to achieve extremely good load balancing, i.e., (amount of work) / (processing capability) is uniform among all nodes. For homogeneous systems, if the workload domain is easily decomposable, then simple uniform partitioning will suffice, and multitudes of studies have been conducted to achieve proper domain decomposition for more complex algorithms. Such load balancing work can be readily be applied even for nodes that are composed of heterogeneous elements, provided that (a) the architecture of the nodes are largely uniform (homogeneous) across the entire machine, and (b) during execution, the codes will be running simultaneous on one of the processors within the node, all at the same time within the machine. Practically all successful `accelerated' supercomputers and their applications, e.g., GPU machines such as Frontier, follow this pattern. 

However, once the nodes are composed of a plethora of customized hardware, and expected to be utilized in a more heterogeneous fashion as in a smartphone, load balancing becomes extremely difficult, and thus weak scaling speedup will flatten quickly, especially in a large parallel system. There have been efforts to alleviate this by creating a task graph of the workload and conduct dynamic load balancing, but have not really achieved success for very large systems, let alone for numerous heterogeneous accelerators.

\begin{comment}
In fact, if strong scaling works very well to the limit and at the same time we increase the # of nodes  n,  then time for the accelerated component will be zero, leading to 1/(1-t) speedup (assuming t was the fraction of the original non-accelerated execution time) while the weak scaling speedup approaches the ideal n asymptotically, so entire speedup ~= n/(1-t). Any perturbation, however, drastically affects this, more so than non-accelerated systems. Let us say that for 99.9% of the nodes acceleration is 90%, but on one node for some reason we achieve only 10%. Then the whole speedup shrinks from 9x to just 1.1x. 
 
Eliminating such a case is very hard, as it requires very good scheduling algorithm based on the prediction on the runtime of each. heterogeneously accelerated component. The simple solutions are either (1) we have homogenous configurations across nodes, and conduct standard domain decompositioning, or (2) the workload is independently divisible such that we can splice up the work into arbitrary small chunks and scheduled on-demand basis like work stealing.
 
Then people try all these asynchronous task graph stuff, but...
\end{comment}

This is why, even for GPU-based machines, not only the node architectures are homogeneous, but also, in any given workload only GPUs or CPUs are used dominantly, but not typically both. Contrastingly, that large parallel program decomposed into a smaller task/dataflow graph and executed on-demand  heterogeneously on a plethora of accelerators is only largely beneficial for small programs on a small machine, but not for HPC where parallelism will continue to increase to exploit weak scaling

The second reason is the increasing difficulty of dark silicon being available in the system to be utilized for heterogeneously specialized hardware, for cost reasons. In the past, dark silicon was projected to be abundant with reduced lithography, thus justifying the ``plethora of accelerators'' view, as they were available for a very low cost. However, with the slowing down of Moore's law, coupled with the high cost of manufacturing due to more advanced fab technologies such as EUV, transistor cost over time is flattening, or may even increase. Thus, the chip cost will become largely proportional to the number of transistors irrespective of the lithography, so every transistor has to contribute to the overall performance improvements in a major fashion, turning dark silicon into expensive unused silicon.

For smartphones, the major cost of the phone is not the SoC but rather in the peripherals such as the screen, camera, flash memory, etc., and the battery life is premium in the cost metric so extra cost incurred by dark silicon may be tolerable. For supercomputers, however, the major cost of the machine is the processors themselves, dominating over 50\% of the overall capital expenditure (CAPEX). So unless the acceleration could benefit some major proportion of the workload, dark silicon would become an intolerable waste. That is why, over generations, accelerators such as GPUs tend to become more general purpose to cover an increasing proportion of the workload, ultimately becoming as general purpose as the CPUs (or GPGPUs).

The third reason is software and productivity. Unless the accelerator usage is extremely easy, e.g., hidden under a set of very simple APIs, expecting the programmers to adopt an arcane programming model is not viable. In fact, this is more serious for HPCs where the market for applications is much smaller than major commodity ecosystems such as smartphones, with a less performance-conscious but extremely large market. Thus, for example, a large consumer-oriented IT company such as Apple can afford to replace a part of its API for a phone with hardware because it will sell more than 100 million iPhones, but not for supercomputers that have a much narrower market and thus do not warrant such investment.

\Questions{%
\MQ{1}~Will extreme heterogeneity happen?
\MQ{2}~Are supercomputer workloads worth extreme specialization?
\MQ{3}~When will we have production supercomputers with more than one accelerator type?
}

\Resonses{\noindent\begin{minipage}{\linewidth}
    \setcounter{figure}{2}
    \renewcommand{\figurename}{Myth}
    \begin{minipage}[tbp]{0.41\textwidth}
        \centering
        \includegraphics[width=\linewidth]{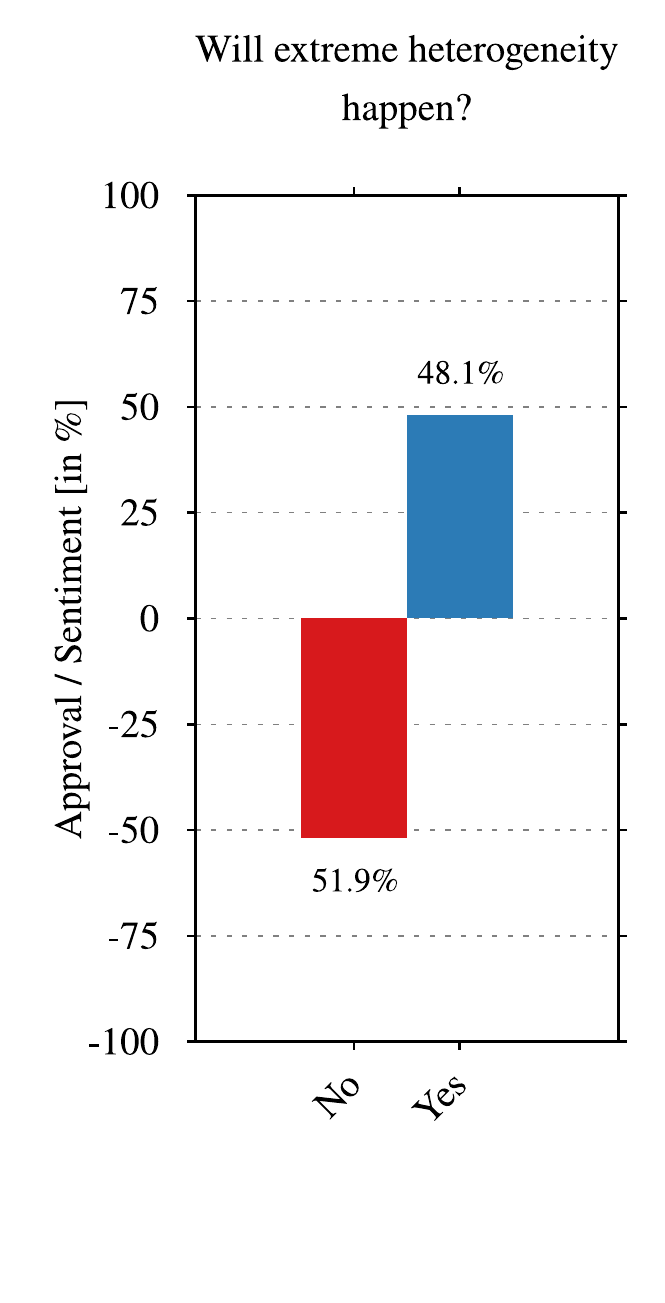}
        \captionof{figure}{Feedback for~\href{https://x.com/thoefler/status/1619809319062749184}{\MQ{1}}}
        \label{fig:myth3}
    \end{minipage}
    \hfill
    \setcounter{figure}{2}
    \begin{minipage}[tbp]{0.533\textwidth}
        \centering
        \includegraphics[width=\linewidth]{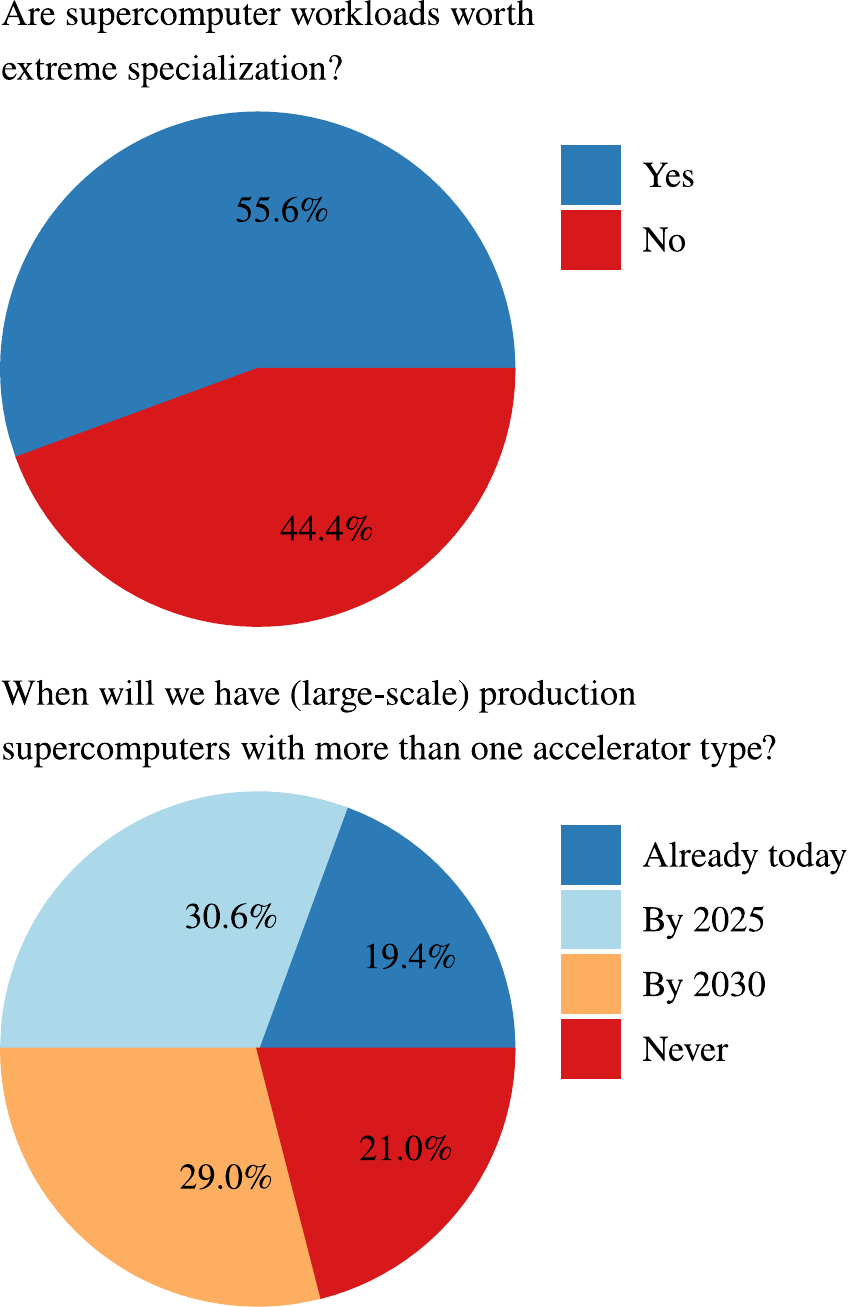}
        \captionof{figure}{Feedback for~\href{https://x.com/thoefler/status/1619809465439776768}{\MQ{2}} \&~\href{https://x.com/thoefler/status/1619809749658390529}{\MQ{3}}}
        \label{fig:myth3b}
    \end{minipage}
\end{minipage}}

\begin{comment}
\textit{We close with the questions:} 
- Despite the difficulties, will extreme heterogeneity happen, in that there would be plausible technologies where load balancing would be achieved in general over multitudes of accelerated hardware and applications realistically?

- Specialization could happen in small forms so as not to require large transistor counts alleviating point 2 above, hidden behind a well-defined API to alleviate point 3, and uniformly applied to the same section of a weak scaling code with predictable and homogeneous runtime across nodes. What would be the acceleration hardware that would satisfy such a requirement, and in which application areas will it be beneficial for?

- Specialization itself may benefit if the machine will be used for a very specific purpose and the return on investment. This has been a case for AI deep learning workload, and possibly on others such as molecular dynamics with Anton. Are there any other major workloads that would benefit, in particular where significant, order of magnitude cost/performance benefit can be attained over general purpose CPUs and/or GPUs?
\end{comment}
\begin{figure*}[tbp]
    \setcounter{figure}{0}
    \centering
    \includegraphics[width=.7\linewidth]{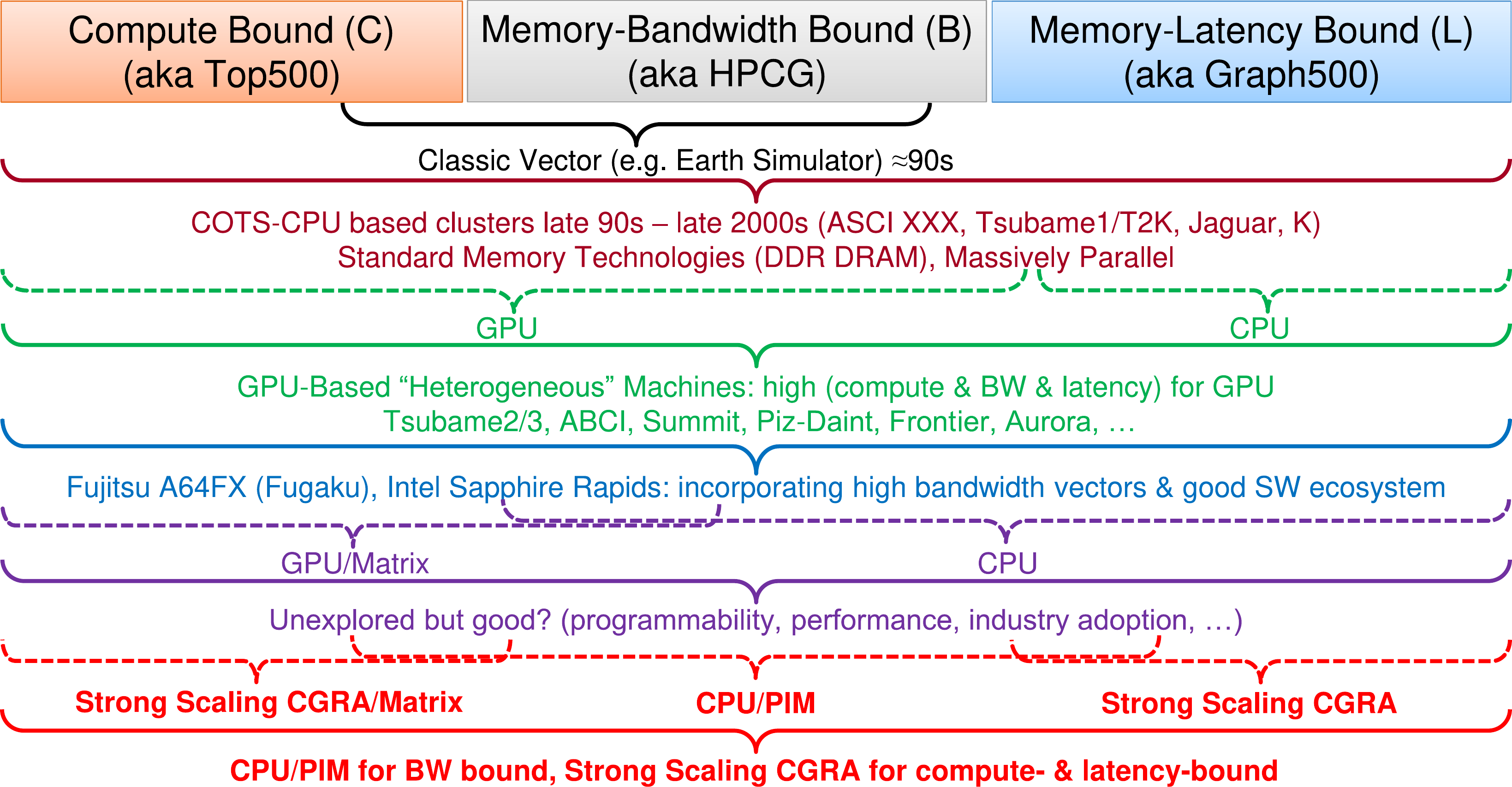}
    \caption{Classification of Compute Kernels and Supercomputing Architecture}
    \label{fig:ref-kernel-classification}
\end{figure*}

\section*{Myth~\nextmyth{sec:accel}: Everything Will Run on Some Accelerator!}%\label{sec:accel}

Related to our previous myth, even if one accepts that there will not be a plethora of accelerators, there could be a few such as GPUs or FPGAs, where the dominant portion of the workload will run. Indeed, for GPU-based machines that would be an assumption, lest the extra investment will not make sense. However, one could question, would some superchip such as GPUs largely replace the CPUs, the latter be degraded to second-class citizens? It is not trivial as it may seem, as such statements are rather dogmatic and not based on candid analysis of the workloads. By proper analysis of the workloads, we may find that CPUs may continue to play a dominant role, with the accelerator being an important but less dominant sidekick.

From the hardware perspective, workloads can be largely divided into three classes, (C) compute-bound, (B) memory bandwidth-bound, and (L) memory latency-bound. Any application will be composed of multiple compute kernels, each one being able to be largely classified into one of the three in Figure~\ref{fig:ref-kernel-classification}. Over time, supercomputer architectures have evolved in an attempt to cover all three in effective ways.

Up until the 90s, special-purpose vector machines such as Cray and NEC SX accelerated largely (B), and (C) to some extent. This was largely due to the dominant workload that was CFD which was largely (B). Then in the 90s, the microprocessor evolution for HPC happened, utilizing the commodity one-chip CPUs which had become very powerful due to high-end applications such as engineering and multimedia needs, starting with workstation/server RISC and then later x86 processors in massively parallel fashion, e.g., DoE ASCI Red. Individual processors were mediocre in performance but attained performance via massive parallelism, exercising weak-scaling, cf.~Myth~\ref{sec:lowpower}.

Then in the late 2000s, although achieving Petascale performance was pioneered with the DoE Roadrunner and Jaguar machines, there was an ambition to achieve exascale by the late 2010s, achieving 1000x scaling in performance in 10 years. The roadblock was power/performance using conventional CPUs. However, by the late 2000s, the GPUs were evolving from their graphics-specific purpose to become general-purpose compute processors, as they were architectural descendants of classical vector processors~\cite{matsuoka_rise_2008}. Different from classical vectors where the floating point performance had been significantly enhanced, motivated by graphical workloads, and when generalized, the GPUs were now covering (C) and (B), while (L) was left for CPUs as the GPU vector pipeline had very long latency. CPUs that facilitated SIMD vector units with high bandwidth memory such as the Intel Xeon Phi and Fujitsu A64FX brought in classical vector properties back into the CPUs, so in a sense, homogeneous systems composed of such chips were not direct reincarnations of simple commodity CPU-based massively parallel machines, but rather, can be more regarded as converging the GPU and CPU properties.

Circa 2022, the top machines are either homogeneously configured heterogeneous CPU-GPU nodes, or `converged' nodes such as RIKEN Fugaku or forthcoming machines with Intel Sapphire Rapids CPUs with HBM. However, this is not the only possible combination, and other configurations have not been properly explored. For example, one could conceive of a machine with the latter configuration, with purpose-built matrix-based accelerators for compute-intensive kernels as a separate chip (or chiplet). In such a machine, the CPU would cover workloads (B) and (L), while the matrix accelerator will cover (C). The benefit of such a machine would be ease of programming of (B) workloads which often involve complex memory access patterns, and thus porting to GPU codes has proven to be challenging.

For further acceleration of (L) workloads, there is a limit to acceleration, such as molecular dynamics that require strong scaling. The best strategy seen for such workloads is fully customized data pipelines such as Anton~\citep{shaw_anton_2008} with hardware design time synthesis. One could almost mimic such customization with cost but make it programmable by FPGAs or CGRAs. Such dataflow customization could also be useful for compute-bound workloads such as DL Transformers, if small matrix engines as special function units can be conjoined in a larger macro dataflow as seen in modern FPGAs and CGRA chips. As such, in such a machine, (B) will be covered by CPUs, while (C) and (L) will be covered by a `strong scaling accelerator'.

As we observe here, we find that we have not even covered the possible configurations of divergence/convergence of processing units, as the only mainstream `accelerated' machines are GPUs with the second property, while other design spaces have not been properly explored.

\Questions{%
\MQ{1}~Will CPUs become pure “servants” to the accelerators?
\MQ{2}~Are accelerators actually more than just better-balanced processors?
\MQ{3}~Will reconfigurable accelerators see a renaissance?
}

\Resonses{\noindent\begin{minipage}{\linewidth}
    \setcounter{figure}{3}
    \renewcommand{\figurename}{Myth}
    \begin{minipage}[tbp]{0.41\textwidth}
        \centering
        \includegraphics[width=\linewidth]{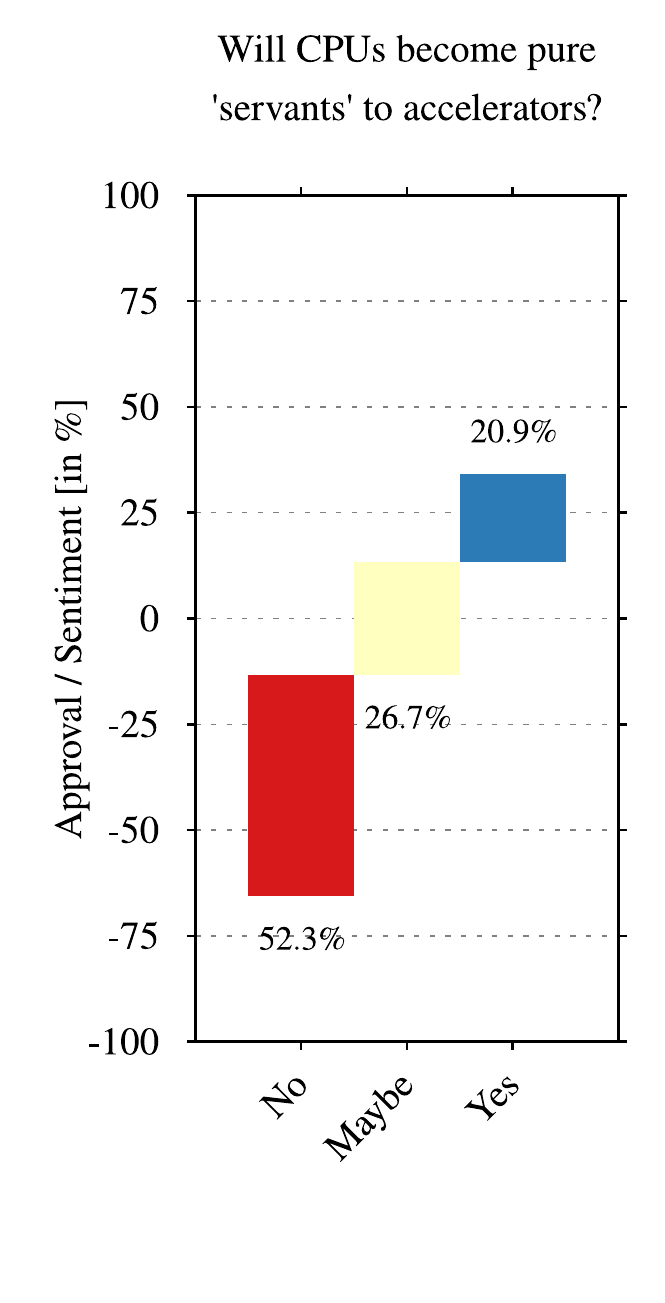}
        \captionof{figure}{Feedback for~\href{https://x.com/thoefler/status/1621414478687440898}{\MQ{1}}}
        \label{fig:myth4}
    \end{minipage}
    \hfill
    \setcounter{figure}{3}
    \begin{minipage}[tbp]{0.513\textwidth}
        \centering
        \includegraphics[width=\linewidth]{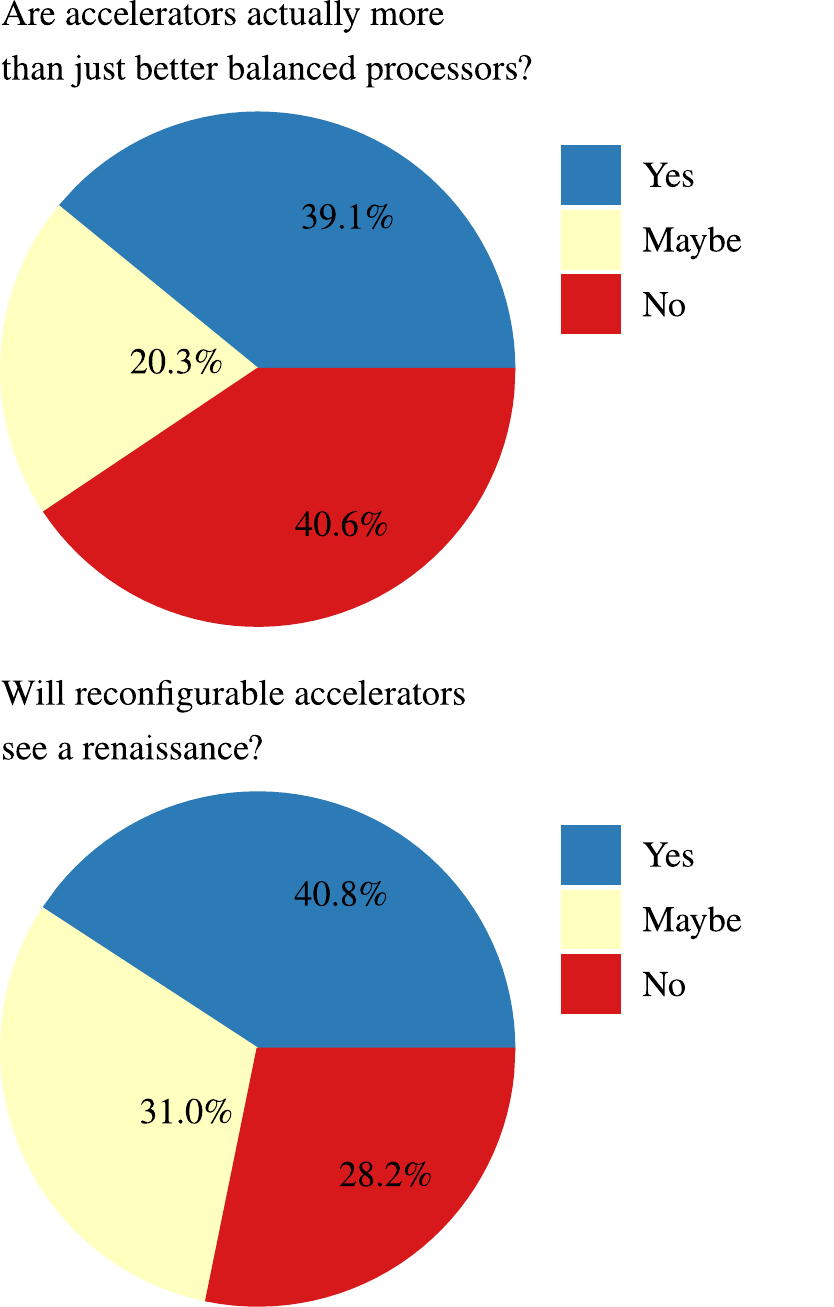}
        \captionof{figure}{Feedback for~\href{https://x.com/thoefler/status/1621414691879768065}{\MQ{2}} \&~\href{https://x.com/thoefler/status/1621414836067356673}{\MQ{3}}}
        \label{fig:myth4b}
    \end{minipage}
\end{minipage}}

\begin{comment}
Moreover, what would be the appropriate programming model for such machines, including the ability to achieve performance advantages as well as portability and productivity of not only new codes but existing codes as well. Will GPUs, their HW and their programming models evolve to accommodate dominant part of the workloads (and thus they become the new CPUs), or will other accelerator configurations evolve, including the converged CPUs (incorporating vector and matrix engines, even some dataflow features) to make GPUs obsolete?

- Are there any other architectural types we have not covered above worthwhile to be considered as an architectural type, and if so, what would be the means of acceleration, and how would they be possibly composed with other processors/hardware?

- Moreover, what would be the appropriate programming model for such machines of such unexplored categories, including the ability to achieve performance advantages as well as portability and productivity of not only new codes but existing codes as well. 

- Will GPUs, their HW and their programming models evolve to accommodate dominant part of the workloads (and thus they become the new CPUs), or will other accelerator configurations evolve, including the converged CPUs (incorporating vector and matrix engines, even some dataflow features) to make even GPUs obsolete?
\end{comment}

\section*{Myth~\nextmyth{sec:fpga}: Reconfigurable Hardware Will Give You 100X Speedup!}%\label{sec:fpga}

In a ``fool me once\ldots’’ fashion, one accelerator, in particular, has taken the HPC community by storm with lofty promises of 100x speedup~\citep{lee_debunking_2010} ever since the first ported matrix-multiplication by~\cite{larsen_fast_2001}. Fueled by NVIDIA's gross margin of over 50\%~\citep{macrotrends_llc_nvidia_2022}, and supported by billions of dollars from US DOE for ECP and similar programs in other parts of the world, the HPC community eventually migrated to a well-designed and broadly adopted GPU/CUDA ecosystem. Consequently, 164 systems of the TOP500 list utilize accelerators from NVIDIA. Nearly two decades later, Fugaku has shown that it only took long vectors and high-bandwidth memory to match GPU performance and energy-efficiency for many workloads. One positive aspect is that much code has been ``modernized'', i.e., rewritten in CUDA or languages and frameworks promising portability to utilize new devices. But the open question is how portable are these modernized codes really? Can they run seamlessly on all new devices?

The global FPGA market was recently valued at about one-third of the global GPU market~\citep{allied_market_research_graphic_2020,allied_market_research_field_2022}. Major chip vendors buying the leading FPGA hardware vendors, AMD acquired Xilinx and Intel bought Altera, respectively, indicates an interest in FPGA integration into future mainstream products. However, so far this has not materialized. Whether FPGA can replace or complement the mainstream GPUs in the HPC and data center market hinges on the questions regarding the cost-to-performance ratio, an existing software ecosystem, and most importantly the productivity of programmers. Unfortunately, we see hurdles in all these areas, which the community and industry might be able to solve with enough time and money. Without offering at least a factor of 10x performance gain at moderate porting costs, ``FPGAs are not a factor in our current planning, because of their unprogrammability''~\citep{sorensen_special_2019}.

Whether reconfigurable logic can replace or amend GPUs as accelerators is interesting. FPGAs will certainly have a harder time due to their high flexibility which comes at a cost. Units built from reconfigurable logic are 10--20x less energy and performance efficient in silicon area. This issue can be addressed by hardening certain blocks, e.g., floating point units, as some FPGA companies do. However, even then, the whole control path would be much less efficient, and it is unclear whether program-driven execution is that much less efficient compared to reconfigurable dataflow. A new line of reconfigurable accelerators as materialized in Xilinx' adaptive compute acceleration platform are similar to coarse-grained reconfigurable arrays (CGRAs) and offer more programmable blocks with a configurable dataflow interconnect. But if now 90\% of the chip are hardened units, are those devices just GPUs with a less mature ecosystem?

\Questions{%
\MQ{1}~Will the HPC community embrace FPGAs as alternatives to GPUs in large-scale production systems?
\MQ{2}~Can the community afford a ``Fool me twice\ldots’’ moment?
\MQ{3}~Will CGRA-style reconfigurable dataflow accelerators take the place of FPGAs to compete?
}

\Resonses{\noindent\begin{minipage}{\linewidth}
    \setcounter{figure}{4}
    \renewcommand{\figurename}{Myth}
    \begin{minipage}[tbp]{0.41\textwidth}
        \centering
        \includegraphics[width=\linewidth]{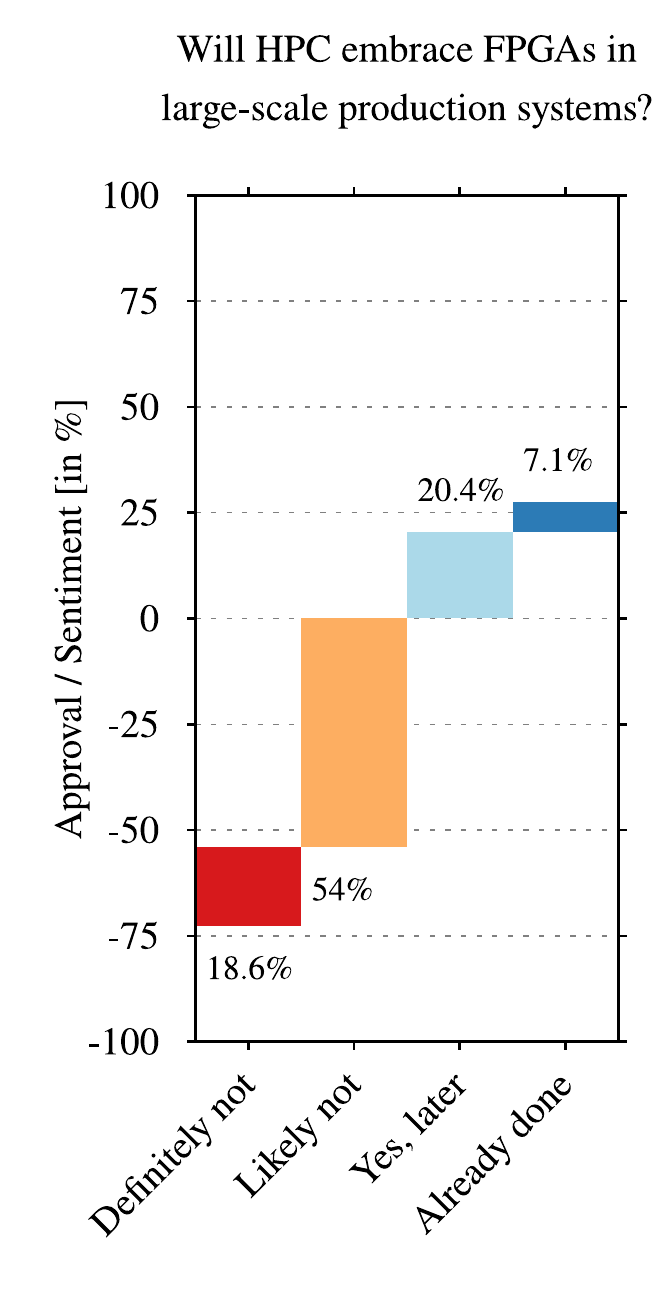}
        \captionof{figure}{Feedback for~\href{https://x.com/thoefler/status/1622533276781912066}{\MQ{1}}}
        \label{fig:myth5}
    \end{minipage}
    \hfill
    \setcounter{figure}{4}
    \begin{minipage}[tbp]{0.567\textwidth}
        \centering
        \includegraphics[width=\linewidth]{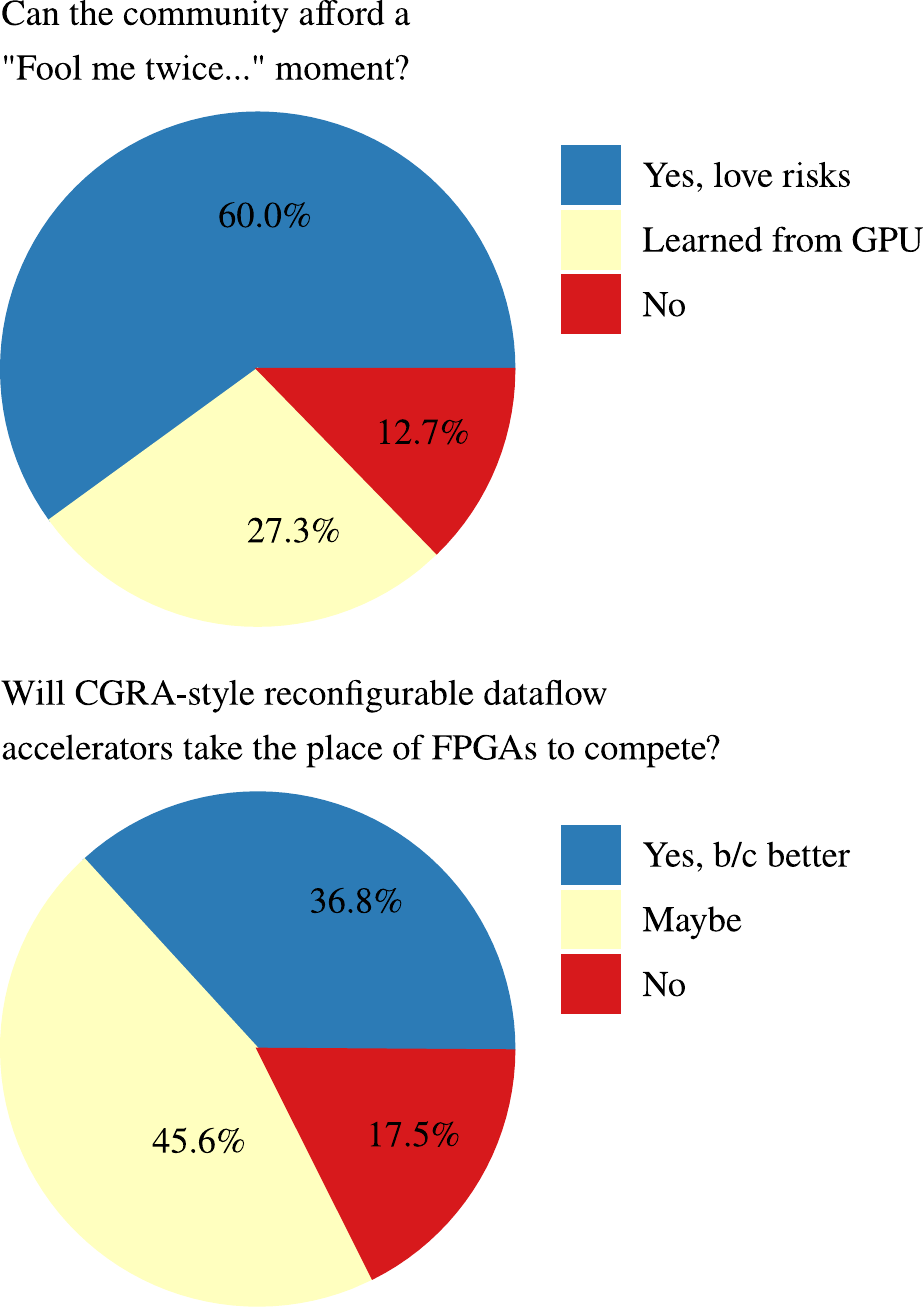}
        \captionof{figure}{Feedback for~\href{https://x.com/thoefler/status/1622533568663519234}{\MQ{2}} \&~\href{https://x.com/thoefler/status/1622533809508868099}{\MQ{3}}}
        \label{fig:myth5b}
    \end{minipage}
\end{minipage}}

\begin{figure*}[tbp]
    \setcounter{figure}{1}
    \centering
    \includegraphics[width=.7\linewidth]{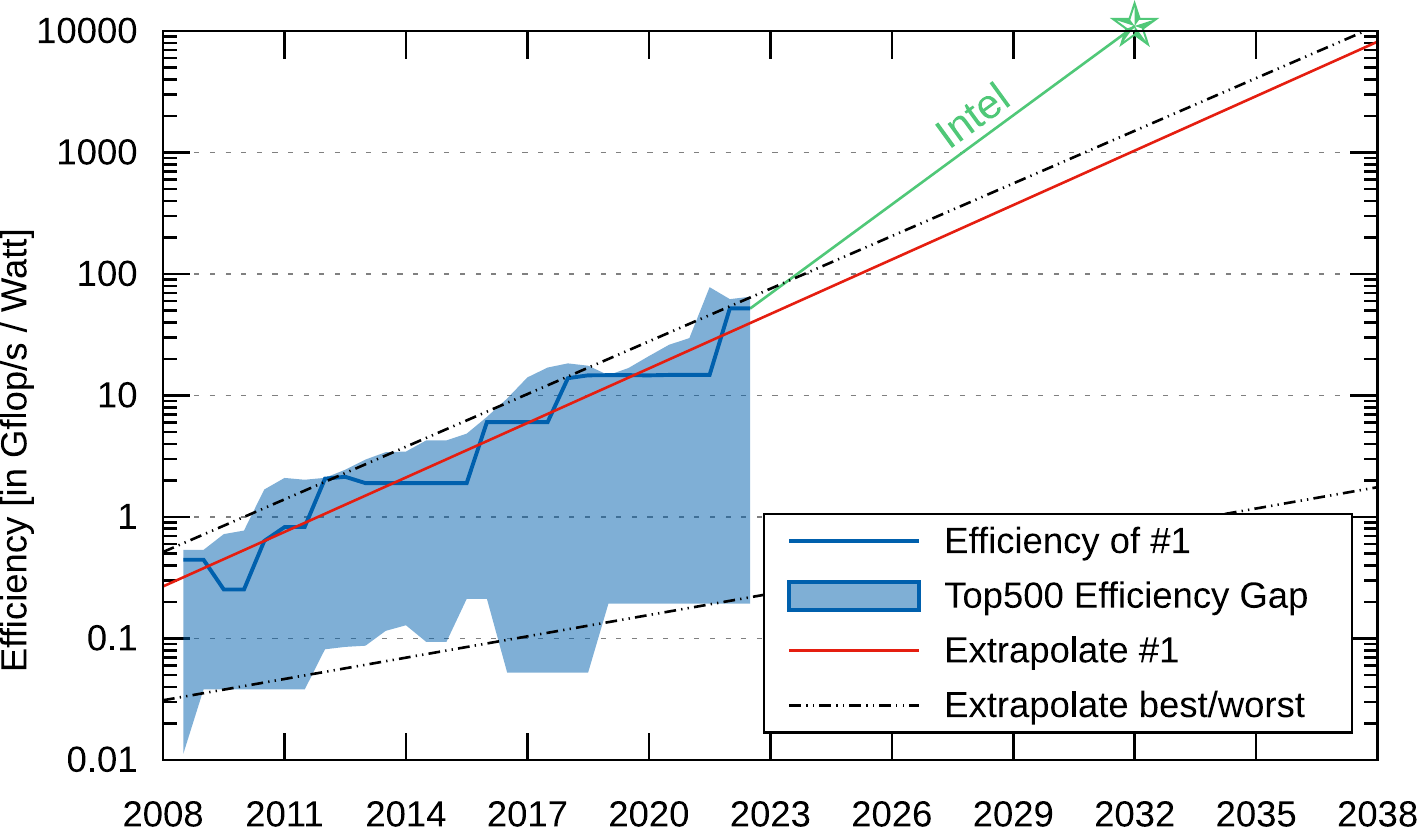}
    \caption{Historical fp64 power efficiency [in $\frac{\text{Gflop/s}}{\text{Watt}}$] extrapolated until 2038 to put Intel's zettaflop/s claims into perspective.}
    \label{fig:gflopswatt}
\end{figure*}

\section*{Myth~\nextmyth{sec:zetta}: We Will Soon Run at Zettascale!}%\label{sec:zetta}

Maybe FPGAs are the way to zettascale. With Aurora still under construction, Intel ignited the debate about zettascale in late 2021. While the HPC community initially smirked at their plans, Intel continued pushing the zettascale agenda, culminating in the latest claims to achieve \unit[1]{zettaflop/s} by the end of the decade~\citep{cutress_interview_2022}. This proposition needs to be addressed, and we try to put their claims into perspective and predict a realistic timeline. Obviously, we cannot rule out that Intel has a secret, revolutionary technology that they plan to commercialize in due time, however, let us not speculate now and instead build on publicly available data.

But first, we have to distinguish the terms. We assume in the following, that (1) ``zettaflop system'' refers to any computer capable of achieving over $10^{21}$ double-precision floating-point operations (``FP64'') per second on the Linpack benchmark; (2) ``zettaop system'' refers to any computer theoretically capable of performing $10^{21}$ operations\footnote{An exact and consistent definition of ``operation'' in this context is still debated in the HPC community.} per second, and (3) ``zettascale system'' denotes any computer executing a scientific application with a sustained performance of over \unit[1]{zettaflop/s} in fp64.

Before we extrapolate, we look at historical trends by~\cite{strohmaier_top500_2022}. The HPC community achieved \unit[1.068]{teraflop/s} with Sandia/Intel’s ASCI Red in the summer of 1997, \unit[1.026]{petaflop/s} with Los Alamos/IBM's Roadrunner in the summer of 2008, and achieved (unofficially) \unit[1.05]{exaflop/s} in spring of 2021 with China's OceanLight system and \unit[1.1]{exaflop/s} with Oak Ridge/HPE's Frontier in summer 2022. Not only do 11 and 13 years lie in between these achievements, respectively, but also multiple megawatts. ASCI Red consumed ``only'' \unit[0.850]{MW}, Roadrunner increased that to \unit[2.35]{MW}, and OceanLight and Frontier now consume \unit[35]{MW} and \unit[21.1]{MW}, respectively. This and Figure~\ref{fig:gflopswatt} show that the energy efficiency of modern chips cannot keep up with the demand for increased computing.

Back to Intel claiming to manage 2x performance improvements year-over-year, which would yield zettaflop/s by 2032---but at a power requirement of the entire system of \unit[50--100]{MW}~\citep{cutress_key_2022}. Hence, this 1,000x in performance comes at the cost of 3--5x in power; and reformulated: the energy efficiency to perform fp64 operations needs to increase by 200--350x, from $\approx$50 to over~\unit[10.000]{$\frac{\text{Gflop/s}}{\text{Watt}}$}. Even under idealized conditions and using Frontier’s Rpeak as the baseline, this goal requires a 125x improvement in 10 years, and all of that while other big players slowly acknowledge the end of practical silicon scaling laws~\citep{white_nvidia_2022}. If we believe the~\cite{ieee_irds_international_2021-1} roadmap, we might gain 5x in power density (optimistically rounded from 4.27x) by 2034 at \unit[7]{\AA} compared to \unit[5]{nm}. This leaves 25x, which we could split into 5x from increased transistor count per chip and 5x from increased node count per system. Can we cool the former, yes~\citep{wu_ultra_2021}, and can we interconnect the latter? Sure, but doing so, at \unit[2.5]{GW}, comes down to the will to invest more than anything else, and without revolutions in memory and interconnect technologies, we might see Linpack transition into memory- or I/O-bound territory, nullifying any computational advances.

On the other hand, a zettaop/s system at \unit[100]{MW} in 2032 is far more likely, since low-precision units (such as tensor cores) can boost the $\frac{\text{op/s}}{\text{Watt}}$ metric, e.g., currently fp16 tensor cores demonstrate an 8x advantage over fp64 vector units. Lowering the precision further from fp16 to 3-bit operands could allow for another 5x improvement~\citep{frantar_gptq_2022}, but only if the industry (and HPC community) sees the need for adding these low-precision units, as we discuss in Myth~\ref{sec:lowprecision}. Considering the above, our more realistic, yet optimistic, timeline for zetta is zettaop/s in 2032 at \unit[50]{MW}, zettaflop/s in 2037 at \unit[200]{MW}, and zettascale by 2038. Can Intel, or anybody else, pull it off before then? Only time will tell.

\Questions{%
\MQ{1}~Will we reach zettaflop/s performance or will fp64 lose relevance before?
\MQ{2}~Will we continue to build more power-hungry supercomputers as we did in the past?
\MQ{3}~Which one will happen first: zettascale, practical quantum advantage, or all internal combustion-based engines cease to be produced?
}

\Resonses{\noindent\begin{minipage}{\linewidth}
    \setcounter{figure}{5}
    \renewcommand{\figurename}{Myth}
    \begin{minipage}[tbp]{0.41\textwidth}
        \centering
        \includegraphics[width=\linewidth]{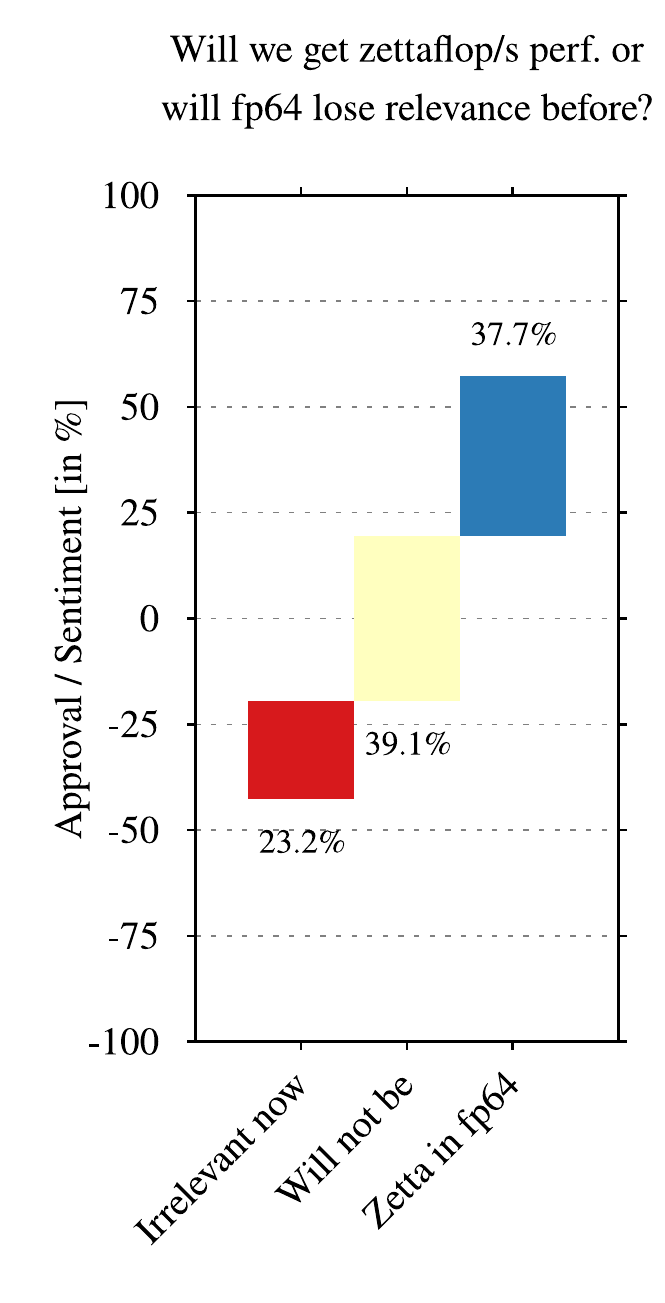}
        \captionof{figure}{Feedback for~\href{https://x.com/thoefler/status/1625039651722473472}{\MQ{1}}}
        \label{fig:myth6}
    \end{minipage}
    \hfill
    \setcounter{figure}{5}
    \begin{minipage}[tbp]{0.567\textwidth}
        \centering
        \includegraphics[width=\linewidth]{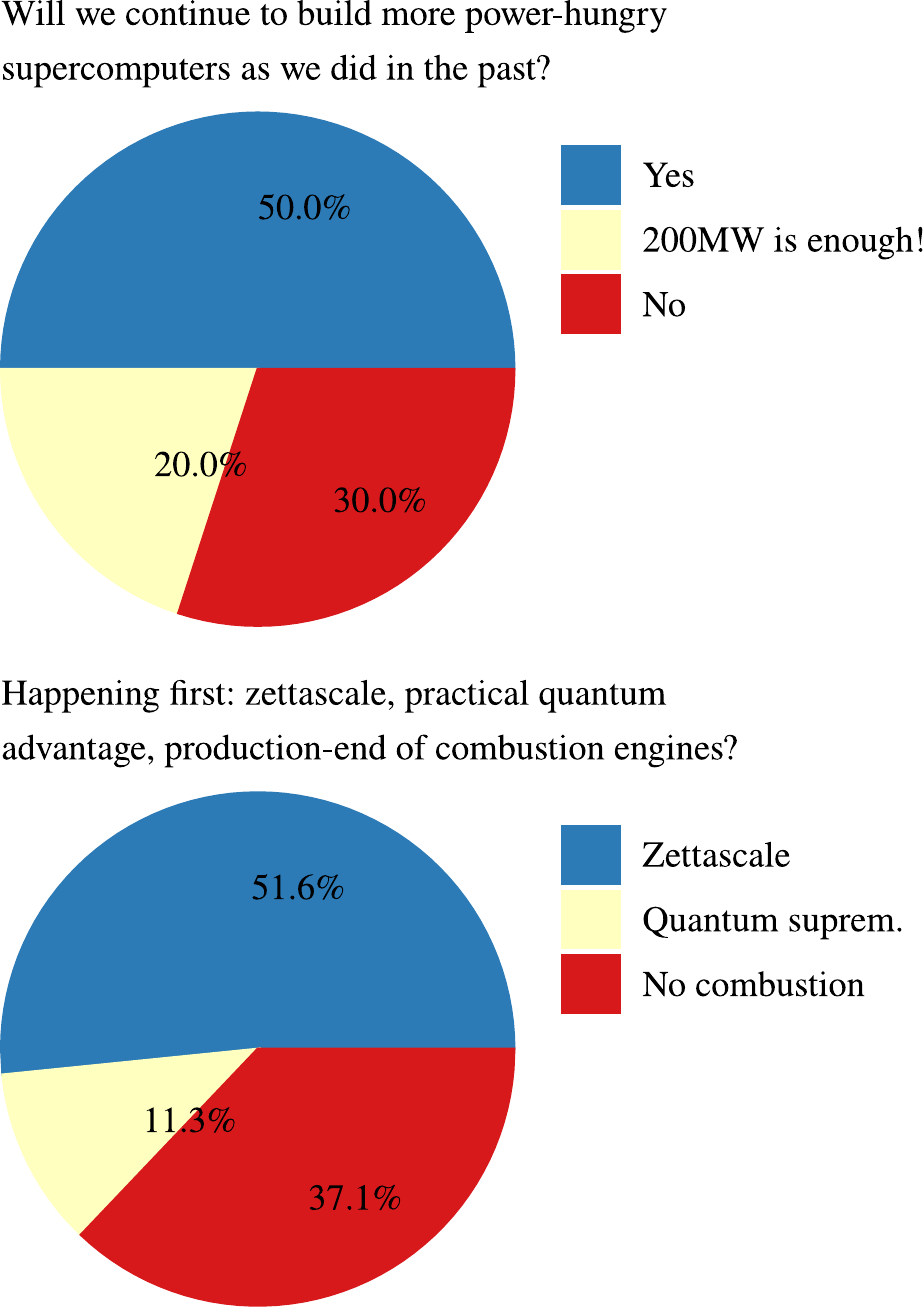}
        \captionof{figure}{Feedback for~\href{https://x.com/thoefler/status/1625039881100578822}{\MQ{2}} \&~\href{https://x.com/thoefler/status/1625040156070670337}{\MQ{3}}}
        \label{fig:myth6b}
    \end{minipage}
\end{minipage}}
\section*{Myth~\nextmyth{sec:resource}: Next-Generation Systems Need More Memory per Core!}%\label{sec:resource}

Before, on the road to peta- and exascale, application scientists continuously raised alarms that the memory per core is decreasing with each new computer generation. This was mainly due to the quick growth in the number of cores, while the performance per core was stagnating. Yet, many workloads can keep those cores utilized with a relatively small working set while staging larger amounts of data remotely and/or recomputing parts. Much of this large memory requirement seemingly turns out to be legacy, and somewhat wasteful, design from times when memory space was abundant compared to other resources. 

Simplistic arguments along the lines of ``we need more of~X'' seem to have a solid tradition in our community. For example, the HPC community spent the first decades hunting more floating-point computations per second. Recently, a demand for larger and faster memory replaced this main goal. The community nearly made a complete 180-degree turn, with~\cite{haus_brave_2021} saying ``computation is free'' and~\cite{ivanov_data_2021} showing ``data movement is all you need''. Some even argue that this turn was taken too late due to the fixation on flop/s. While this was all true at the time, the general discussion should really be about the intricate relation between the application requirements and the system capabilities in terms of balance, i.e., the ratio between the different resources such as memory size/bandwidth and compute~\citep{czechowski_balance_2011}. 

These ratios usually shift with chip technology and architectural choices. For example, Moore's law drove the costs for compute on chip down over decades, but off-chip communication was limited by Rent’s rule. This led to the recent data movement crisis. Newly emerging optical off-chip connectivity, see Myth~\ref{sec:sipho}, as well as 3D integrated memory~\citep{domke_at_2022} shifts the balance again and may alleviate many of these aspects, at least at the scale of a single chip. It seems key to understand the malleability of applications, i.e., which resources can be traded for which other resources (e.g., memory capacity for computation bandwidth using recomputation or caching as techniques). Here, specifically, I/O complexity analysis is a tool to deeply understand this trade-off. Once all trade-offs are understood, requirements models~\citep{calotoiu_lightweight_2018} could be used to fix trade-offs into designs. These models could then inform architectural choices as well as hardware developments.

One area to highlight in this context is embedded design, where such trade-offs have long been used to build real systems due to resource scarcity (e.g., battery). While those designs were initially limited to very narrow application domains (e.g., radio signal, audio, or video processing), embedded devices have recently been expanded towards more diverse workloads (``apps''). We believe that HPC can learn from this field by defining clear system design methodologies based on a solid combination of empirical and analytical modeling. More particularly, systems design in HPC can benefit from the embedded systems' doctrine of accounting for over-engineering just as one accounts for under-engineering.

\Questions{%
\MQ{1}~When will the current ``data movement'' focus end?
\MQ{2}~What will be the next bottleneck resource?
\MQ{3}~Will our community be able to adopt a performance modeling discipline to discuss bottlenecks scientifically?
}

\Resonses{\noindent\begin{minipage}{\linewidth}
    \setcounter{figure}{6}
    \renewcommand{\figurename}{Myth}
    \begin{minipage}[tbp]{0.41\textwidth}
        \centering
        \includegraphics[width=\linewidth]{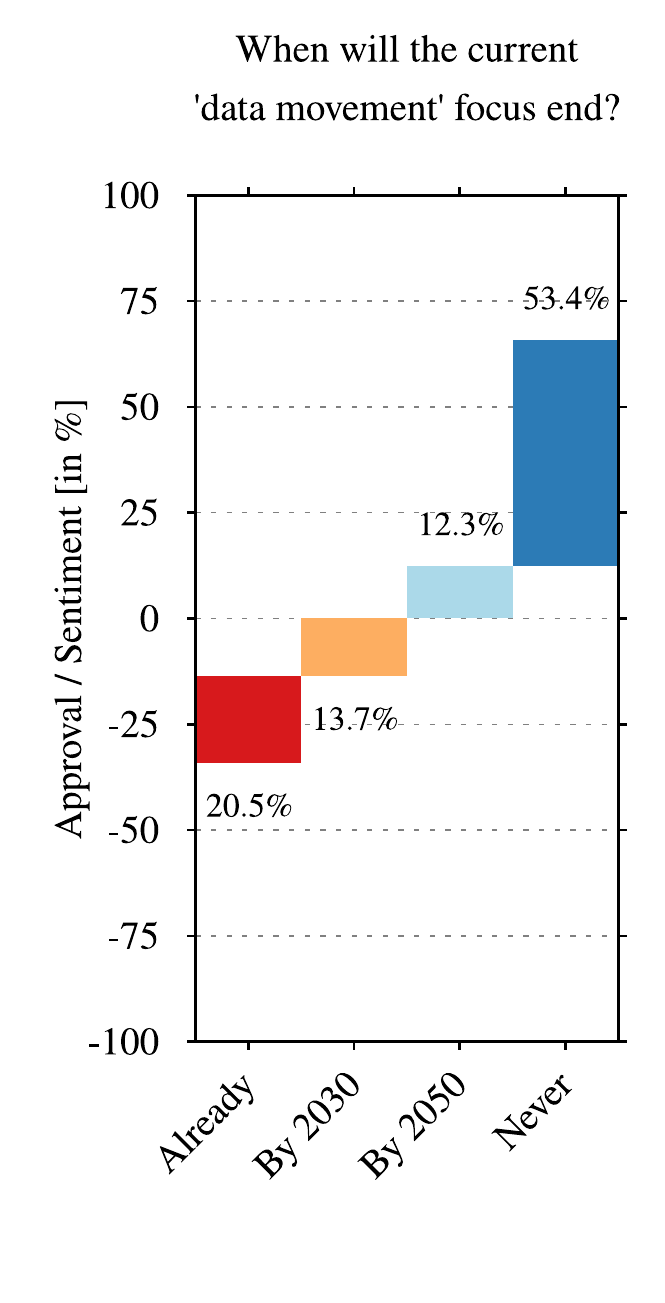}
        \captionof{figure}{Feedback for~\href{https://x.com/thoefler/status/1626494590948704256}{\MQ{1}}}
        \label{fig:myth7}
    \end{minipage}
    \hfill
    \setcounter{figure}{6}
    \begin{minipage}[tbp]{0.560\textwidth}
        \centering
        \includegraphics[width=\linewidth]{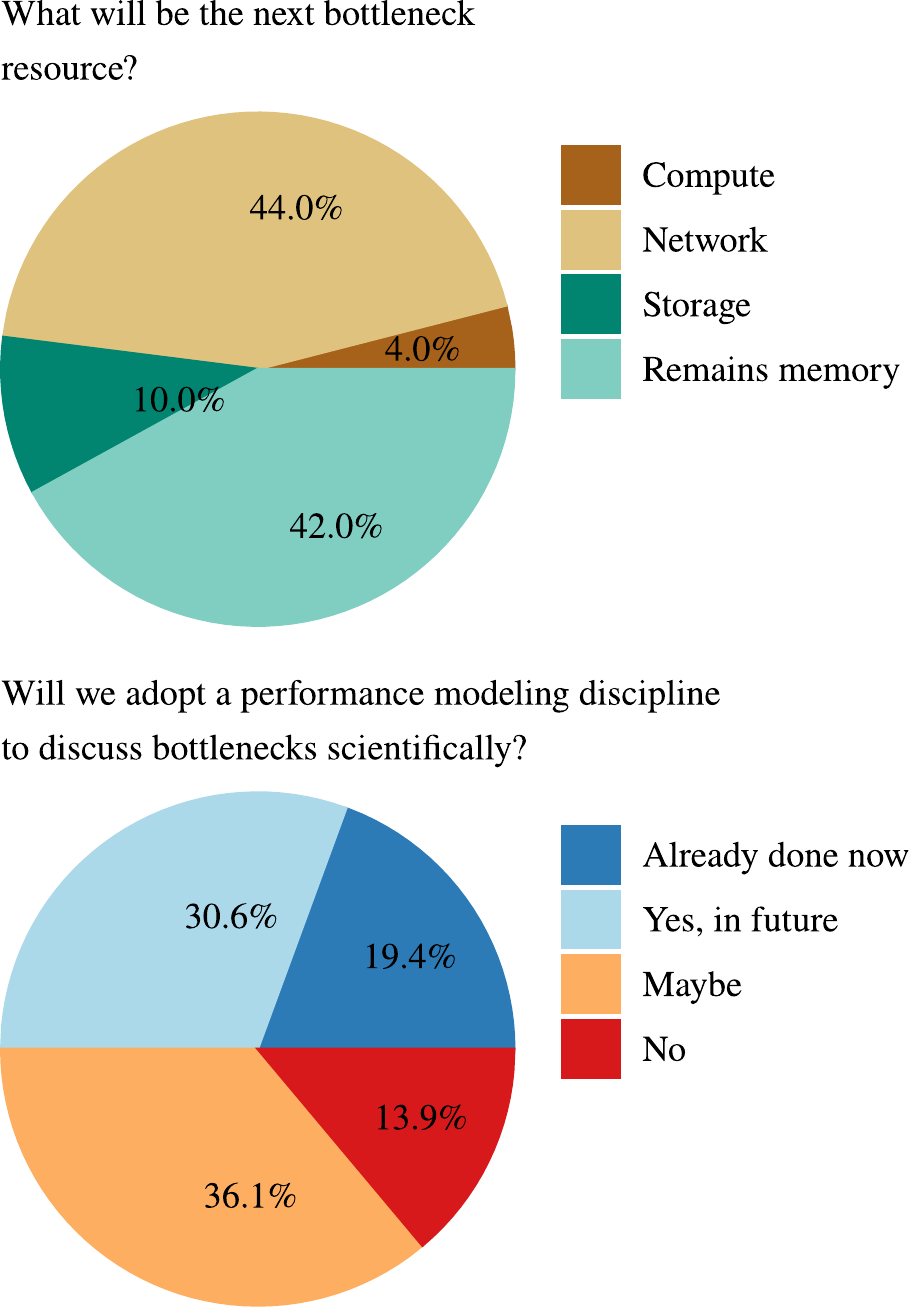}
        \captionof{figure}{Feedback for~\href{https://x.com/thoefler/status/1626494861804204039}{\MQ{2}} \&~\href{https://x.com/thoefler/status/1626495052116549633}{\MQ{3}}}
        \label{fig:myth7b}
    \end{minipage}
\end{minipage}}
\section*{Myth~\nextmyth{sec:sipho}: Everything Will Be Disaggregated!}%\label{sec:sipho}

To stop the waste of memory resources, the academic community is advancing on the Silicon Photonics front~\citep{abali_disaggregated_2015,gonzalez_optically_2022} and industry is pursuing scale-out technologies~\citep{li_pond_2022}, such as Compute Express Link\textsuperscript{TM} (CXL), a cache-coherent interconnect for data centers. But a few players seem to push the idea over the edge with their plans to disaggregate everything~\citep{ntt_rd_what_2020,shan_towards_2022}. As~\cite{gonzalez_optically_2022} stated: ``An optical interconnect is more appealing than an electrical interconnect for memory disaggregation due to three properties: its (1) high bandwidth density significantly reduces the number of IO lanes, (2) power consumption and crosstalk do not increase with distance, and (3) propagation loss is low.'' However, several barriers remain before we can fully replace copper-based interconnects in our supercomputers.

Generally, we see two remaining challenges for the broad adoption of Silicon Photonics and all-optical interconnects: low-cost manufacturing and optical switching. The former is obvious, because after all, the data center and HPC communities rely on inexpensive components to optimize the overall system performance-to-cost ratio. The latter challenge is less obvious for the uninitiated. Current electrically switched networks can operate in ``packet switching'' mode to effectively lower the observable latency and utilize the available link bandwidth. The alternative to this mode is ``circuit-switching’’ and it was abandoned by the HPC community long ago in favor of packet-switching. However, without (cost-)effective means to buffer light, process photon headers in-flight, or reverting to electric switches with expensive optical-electrical-optical conversions, we would have to resort to circuit-switching~\citep{bergman_silicon_2022} with all the inherent deficiencies: complex traffic steering calculations, switching delays, latency increase due to lack of available paths, under-utilization of links, just to name some. 

For HPC, an extensive or extreme disaggregation yields another challenge, specifically the speed of light. Photons travel at a maximum speed of \unit[3.3]{ns/m} in hollow fibers (or slower in other transport media). This is equivalent to a level-2 cache access of a modern CPU, but does not yet include the disaggregation overhead, such as from the CXL protocol itself, switching, or optical-electrical conversions at the endpoints. At \unit[3--4]{m} distance, the photon travel time alone exceeds the first-word access latency of modern DDR memory. Therefore, if main memory would be disaggregated beyond rack boundaries, it will become noticeable for memory-latency sensitive applications, cf.~Myth~\ref{sec:accel}. The more sensible solution for future HPC systems, in line with Myth~\ref{sec:resource}, is to use smaller node-local memory configurations (e.g., HBM3) paired with rack-local, CXL-based memory pools, if the capacity- and performance-to-cost ratios of the memory pool (plus needed interconnect) can outperform node-local SSD solutions.

\Questions{%
\MQ{1}~Will CXL be deployed widely in HPC?
\MQ{2}~Will large-scale supercomputers be disaggregated beyond rack-scale?
\MQ{3}~Should we disaggregate main memory?
}

\Resonses{\noindent\begin{minipage}{\linewidth}
    \setcounter{figure}{7}
    \renewcommand{\figurename}{Myth}
    \begin{minipage}[tbp]{0.41\textwidth}
        \centering
        \includegraphics[width=\linewidth]{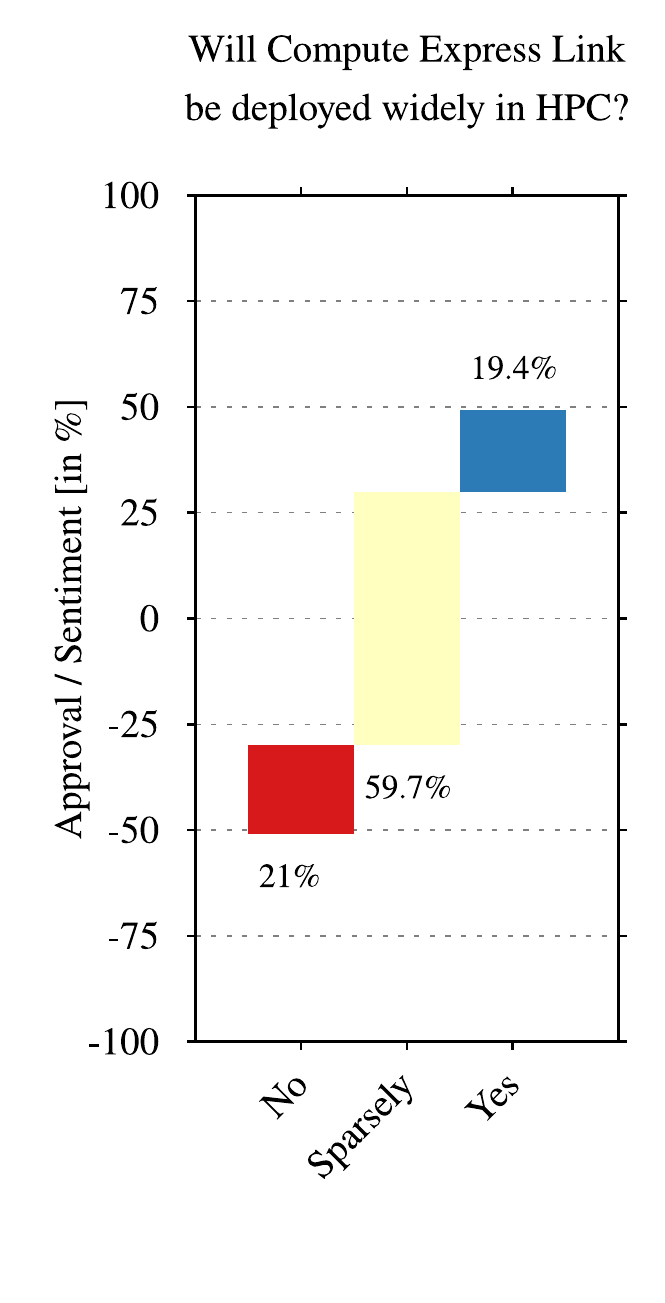}
        \captionof{figure}{Feedback for~\href{https://x.com/thoefler/status/1627604487899693061}{\MQ{1}}}
        \label{fig:myth8}
    \end{minipage}
    \hfill
    \setcounter{figure}{7}
    \begin{minipage}[tbp]{0.504\textwidth}
        \centering
        \includegraphics[width=\linewidth]{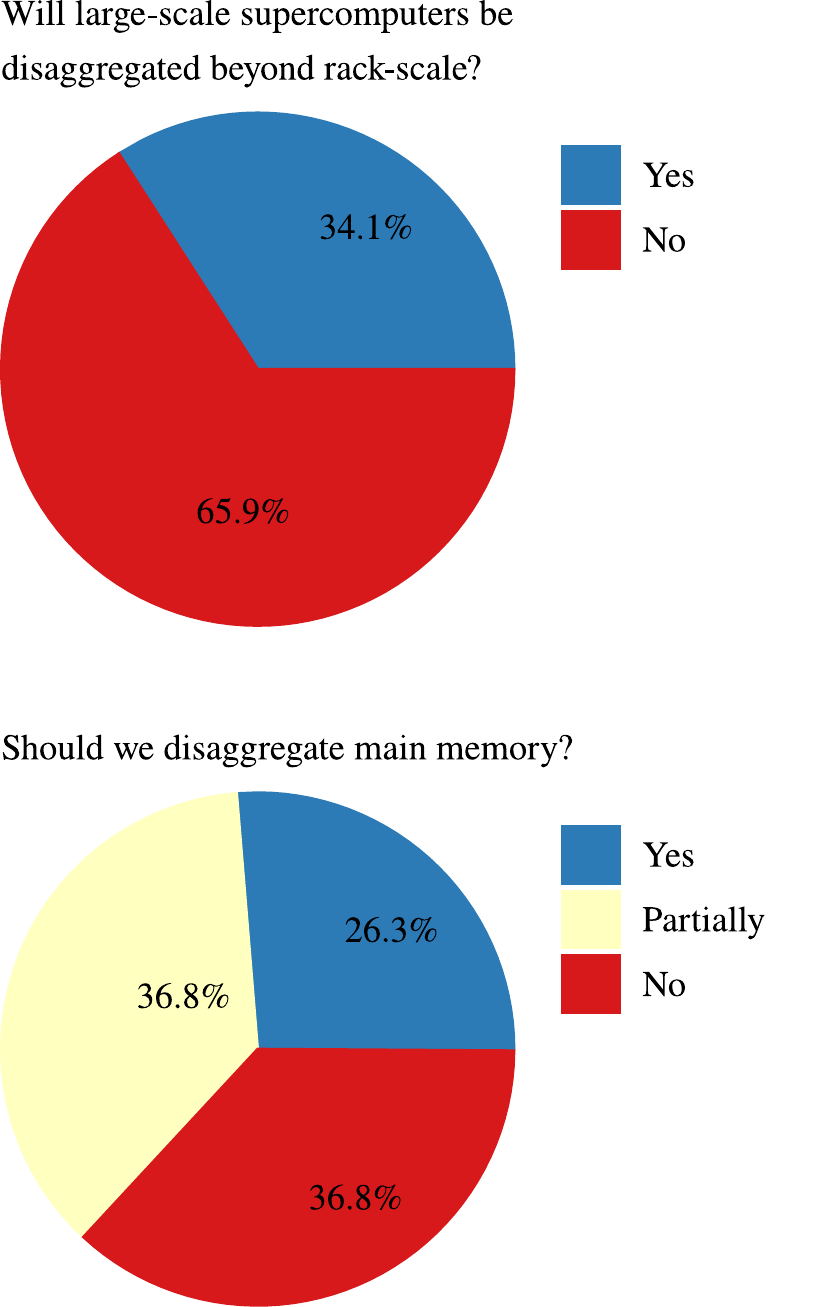}
        \captionof{figure}{Feedback for~\href{https://x.com/thoefler/status/1627604576927973376}{\MQ{2}} \&~\href{https://x.com/thoefler/status/1627604721602002945}{\MQ{3}}}
        \label{fig:myth8b}
    \end{minipage}
\end{minipage}}

%\textcolor{red}{IBM reached out and like to add https://arxiv.org/abs/1503.01416}

\begin{figure*}[tbp]
    \setcounter{figure}{2}
    \centering
    \includegraphics[width=.7\linewidth]{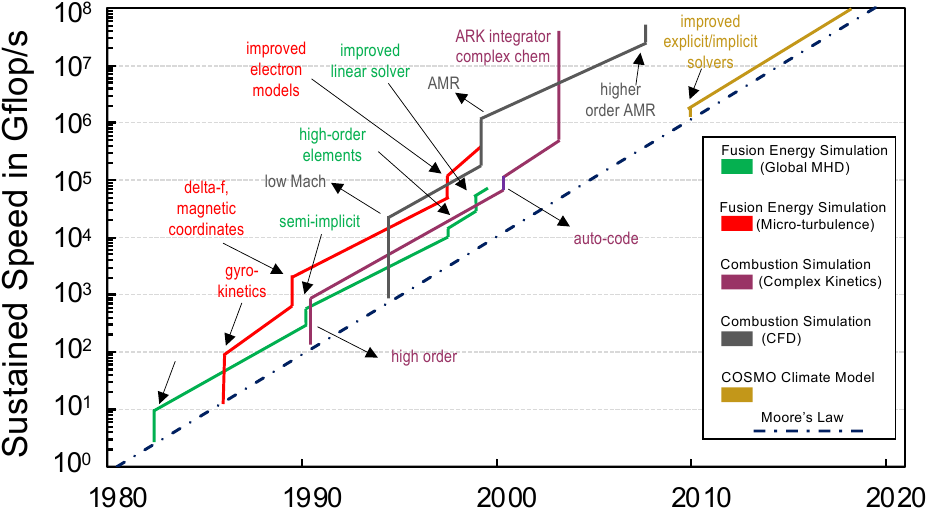}
    \caption{Examples of ``Algorithmic Moore's Law'' for different areas in HPC; Fusion energy and combustion simulation data by~\cite{keyes_efficient_2022} and climate simulation data by~\cite{schulthess_exascale_2016}}
    \label{fig:mooreapps}
\end{figure*}

\section*{Myth~\nextmyth{sec:mooreapps}: Applications Continue to Improve, Even on Stagnating Hardware!}%\label{sec:mooreapps}

Modernizing hardware, with Silicon Photonics, Tensor Cores, or simply shrinking transistors, has too long been the primary method of accelerating legacy software. More than half of this improvement was based on Moore’s law and its observation that transistors will continue to become smaller every few years (originally 18 months). The remaining hardware improvements came from architectural innovations, such as deeper cache hierarchies, the migration to more specialized architectures (e.g., GPUs), or the utilization of larger and wider vector-units (SIMD), as well as scaling the HPC systems up by giving them more processors and cores.

Unfortunately, we are no longer seeing the consistent technology scaling that Moore observed. Consequently, in the so-called Post-Moore era, the ``performance road'' forks three-ways, yielding the following options: (1) architectural innovations will attempt to close the performance gap, and an explosion of diverging architectures tailored for specific science domains will emerge, or (2) alternative materials and technologies (e.g., non-CMOS technologies) that allow the spirit of Moore’s law to continue for a foreseeable future, or (3) we abandon the von-Neumann paradigm together and move to a neuromorphic or quantum-like computer (which, in time, might or might not become practical as discussed in Myth~\ref{sec:quantum}). One major aspect that reflects the uncertainty about the future is the initiatives of unprecedented scale: CHIPS act in the US and similar initiatives in other countries in the order of 100s Billion USD, quantum computing initiatives in the order of 10s Billion USD, etc.

But one point that is often overlooked is that algorithmic improvements in HPC (dubbed as ``Algorithmic Moore's Law'' by~\cite{keyes_efficient_2022}) have over time provided exponential improvement in key areas of HPC, see Figure~\ref{fig:mooreapps}. Similar reports attribute a significant portion of the performance improvement in many legacy codes to be from numerical solvers, algorithms, low-precision numerics, system software, etc~\cite{schulthess_exascale_2016}. However, we have to be cautious that---just as hardware improvements have physics and engineering limits---the ``Algorithmic Moore's Law'' also has its limits: numerical stability, hitting asymptotic limits, etc. That being said, those limits might not be as clear and quantifiable as the limits on hardware. That is because even if one numerical method hits its limit, domain experts can often reduce/pre-condition their problem to another numerical method that is more efficient. Yet, once a linear-time algorithm is found, there remains little room for improvement.

\Questions{%
\MQ{1}~As the performance improvements from hardware technologies drop, should the HPC community dramatically increase the investment in software?
\MQ{2}~Will the ``Algorithmic Moore's Law'' end soon as well?
\MQ{3}~To what extent is the HPC community willing to refactor/rewrite legacy codebases when/if hardware stagnates?
}

\Resonses{\noindent\begin{minipage}{\linewidth}
    \setcounter{figure}{8}
    \renewcommand{\figurename}{Myth}
    \begin{minipage}[tbp]{0.41\textwidth}
        \centering
        \includegraphics[width=\linewidth]{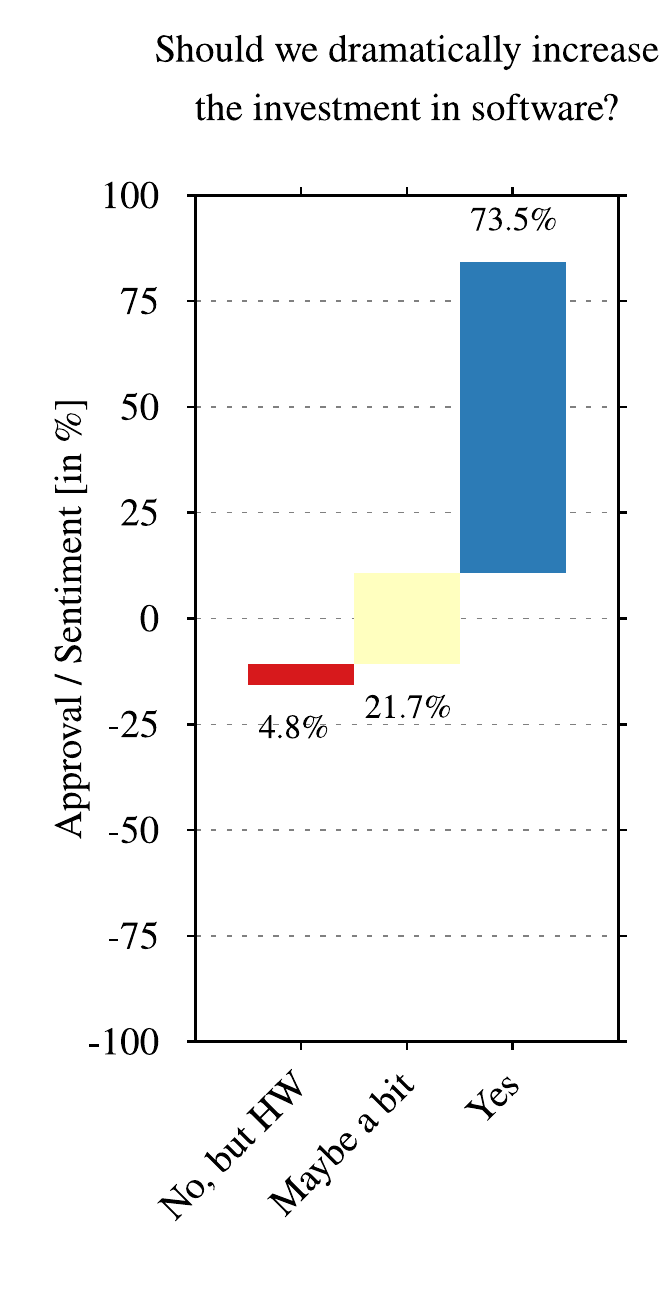}
        \captionof{figure}{Feedback for~\href{https://x.com/thoefler/status/1629032862170112001}{\MQ{1}}}
        \label{fig:myth9}
    \end{minipage}
    \hfill
    \setcounter{figure}{8}
    \begin{minipage}[tbp]{0.535\textwidth}
        \centering
        \includegraphics[width=\linewidth]{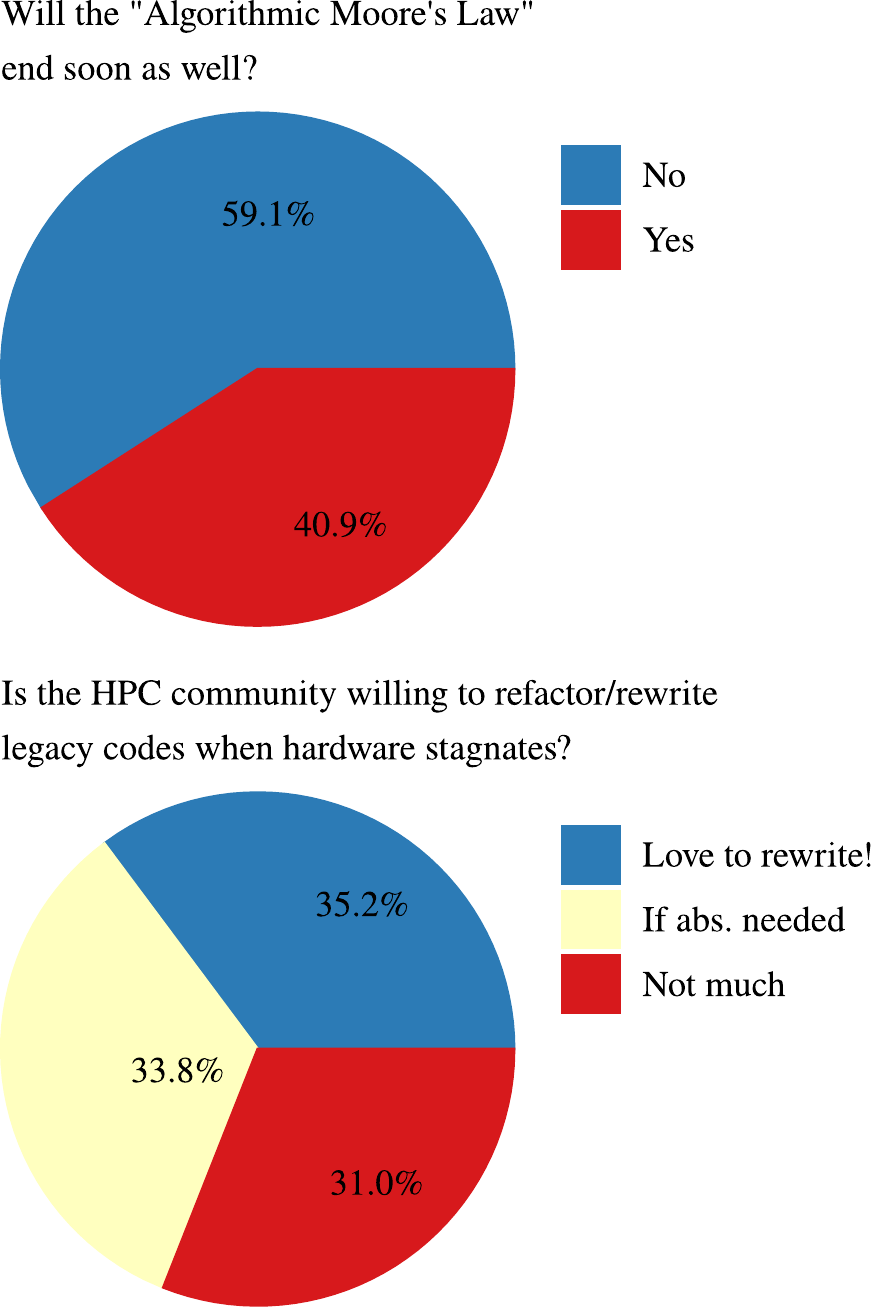}
        \captionof{figure}{Feedback for~\href{https://x.com/thoefler/status/1629033038016290816}{\MQ{2}} \&~\href{https://x.com/thoefler/status/1629033349439160321}{\MQ{3}}}
        \label{fig:myth9b}
    \end{minipage}
\end{minipage}}
%\section*{Myth~\nextmyth{sec:fortran}: Every DSL Dies in Practice (And Survivors Become General Purpose)!}%\label{sec:fortran}
\section*{Myth~\nextmyth{sec:fortran}: Fortran Is Dead, Long Live the DSL!}%\label{sec:fortran}

Applications might have limits, but what about languages? How often have we heard ``Fortran is dead, long live X''? Slogans like this have been resonating in the community for nearly 40 years~\citep{post_real_1982}. X has been everything from C to C++, and more recently Python or Domain-Specific Languages (DSLs). Yet, Fortran remains in wide use in important communities such as weather and climate, even for newly written codes. Other languages, such as COBOL, were indeed replaced with more modern alternatives such as Java. Why is this? Are some parts of our community just stubborn to follow the youngsters? Or are old languages not necessarily bad for the task? Indeed, Fortran is a very well-designed language for its purpose of expressing mathematical programs at the highest performance. It seems hard to replace it with C or other languages and outperform it or even achieve the same baseline. This may be due to the highly optimized Fortran compilers or the limited language features (e.g., no pointer aliasing) that enable more powerful optimizations. 

Fortran and other general-purpose languages remain competitive with many DSLs on CPUs~\citep{ben-nun_productive_2022} and are recently also adopted to GPUs, albeit often less elegant. General-purpose portability approaches such as SYCL~\citep{keryell_khronos_2015}, also powering Intel's oneAPI, or OpenMP provide flexibility as well as some portability across devices. High-productivity general-purpose languages are hard to accelerate in practice. For example, Python's flexibility (e.g., monkey patching and flexible typing) disables many static optimizations. However, when restricting the syntax to high-performance Python (much of NumPy), then optimizations become simpler~\citep{ziogas_productivity_2021}. Any language becomes more complex over time---Fortran 66 evolved into the complex Fortran 2018 language standard. Similar trends affect DSLs that are widening their scope over time. Do we require this generality? If yes, then DSLs are doomed to fail, or they morph into general-purpose languages.

Another argument is that the lower levels usually remain C/C++ and programmers interested in the highest performance are often happy to dig into the lower levels. Then the question remains---where should the portability layer be located? At a (virtualized) Instruction Set Architecture (ISA) as in LLVM's IR~\citep{lattner_llvm_2004}, some lower-level language such as C/C++ as in SYCL/oneAPI, or even dataflow graph representations as in DaCe~\citep{ben-nun_stateful_2019}?

\Questions{%
\MQ{1}~When will programmers stop using Fortran for new applications?
\MQ{2}~Will we ever have more application codes written in DSLs than general-purpose languages?
\MQ{3}~What will be the next big DSL?
}

\Resonses{\noindent\begin{minipage}{\linewidth}
    \setcounter{figure}{9}
    \renewcommand{\figurename}{Myth}
    \begin{minipage}[tbp]{0.422\textwidth}
        \centering
        \includegraphics[width=\linewidth]{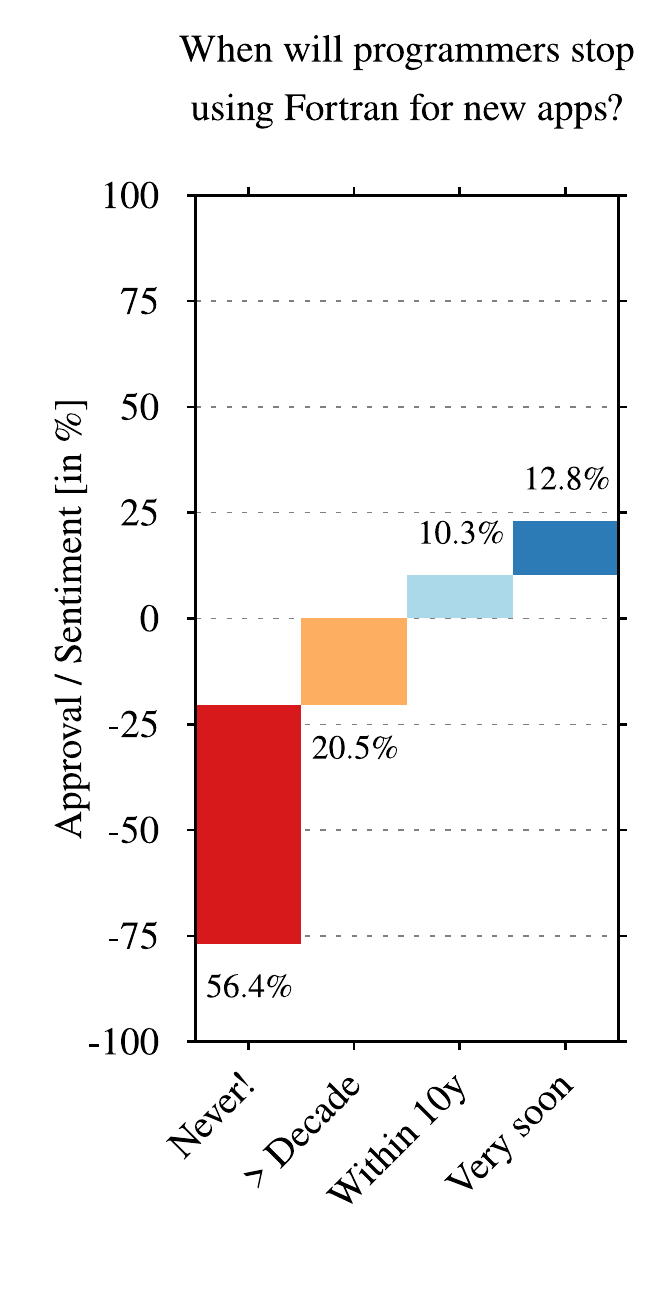}
        \captionof{figure}{Feedback for~\href{https://x.com/thoefler/status/1632994040034603008}{\MQ{1}}}
        \label{fig:myth10}
    \end{minipage}
    \hfill
    \setcounter{figure}{9}
    \begin{minipage}[tbp]{0.561\textwidth}
        \centering
        \includegraphics[width=\linewidth]{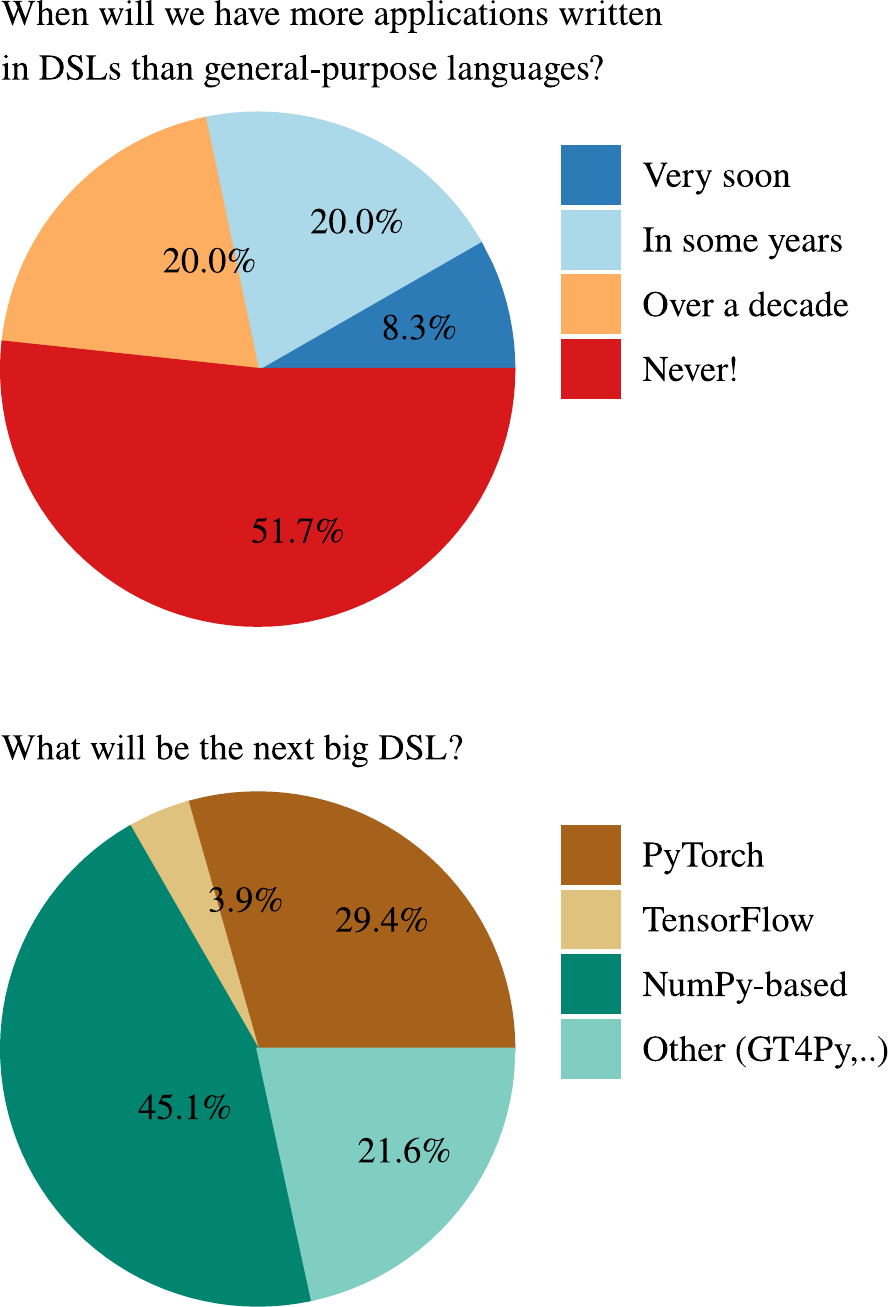}
        \captionof{figure}{Feedback for~\href{https://x.com/thoefler/status/1632994310479048704}{\MQ{2}} \&~\href{https://x.com/thoefler/status/1632994955537838081}{\MQ{3}}}
        \label{fig:myth10b}
    \end{minipage}
\end{minipage}}
\section*{Myth~\nextmyth{sec:lowprecision}: HPC Will Pivot to Low or Mixed Precision!}%\label{sec:lowprecision}

A high-performance language is nothing without proper data types, but high-precision operations such as fp64 come at a significant cost in terms of silicon area, energy, and speed, according to Myth~\ref{sec:zetta}. Lowering this precision can save costs but may reduce the accuracy of the results and, in the worst case, break the application (e.g., convergence). But there is more to this trade-off: what if a more clever implementation could maintain convergence properties of high-precision numerics, while enjoying the computational efficiency of lower precision? One common trick is using mixed precision on the algorithmic level, for example, using low precision for individual particles and only using high precision for aggregated values~\citep{kutzner_more_2019}. Some processors offer mixed precision tricks at the hardware level in the form of instructions with low-precision inputs but higher precision accumulations.
%The later approach particularly shines with systolic arrays \textbf{FIXME (explain why?)} And is also ideal from a productivity standpoint in a sense that this approach requires small to none code modification. 

There is however more to reduced precision than using fewer bits---the question is how to optimally distribute bits between mantissa and exponent~\citep{tesla_inc_whitepaper_2021}, or even if to use an entirely different (not IEEE-754) way to represent numbers~\citep{gustafson_beating_2017}. The story of reduced precision in AI hardware is quite telling: In the early days of the field, predominantly the IEEE fp32 format was used, but knowing that in deep neural nets, the weights and activations are typically distributed on a small range of values, researchers began to explore the fp16 format. Soon the Pascal generation of GPUs with fp16 performance---at a factor of two compared to fp32 was released---and the magic did not happen by itself. Exploding and vanishing gradients, outlier weights, etc., made training large deep neural nets require extra effort to stabilize (incurring corresponding overhead) or just did not converge at all. The next generation of devices came with bfloat16 format---same 16 bits, but more bits allocated to the range, less for the precision. It worked better, but still once in a while, a model would collapse. Finally, the recent generation of GPUs came with a 19-bit numeric format, misleadingly called TensorFloat-32. So far it seems to be at the sweet spot for artificial intelligence workloads---allowing for noticeably faster arithmetics than fp32 while maintaining enough numeric stability for the models to reliably converge without extra programming effort. 

Now that mixed precision is a de facto standard in the AI domain, more hardware support is being implemented. So far, there is no general clarity on the limits---how few bits can we get away with in different HPC areas? The following factors in particular are important to consider as we move forward. A fully transparent solution for the problem is to simulate higher precision using low precision operations, e.g., as shown by~\cite{ootomo_recovering_2022}. Our Myth~\ref{sec:accel}'s memory-bound problems in particular are good candidates for exploiting ``simulated'' high precision, since the overhead can be masked by data transfers. It is not clear however if this incurred overhead is an acceptable price that HPC agrees to pay for remaining in higher precision. A less transparent method is to approach the problem as a precision auto-tuning task by adapting the precision to a minimum while bounding the error, e.g., as demonstrated by~\cite{menon_adapt_2018}. One main limitation of that method is the reliance on automatic differentiation (AD) to track error propagation, which is not practical for large codebases. Finally, the least transparent approach requires domain experts in HPC to study the numerical stability of solvers to identify, on a case-by-case basis, the susceptibility of solvers to lower/mixed precision. While this approach is viable for solvers that are wrapped in libraries to be consumed by HPC domain experts, it is unclear whether domain experts writing their own solvers (common in HPC) would be willing to take on this burden.

\Questions{%
\MQ{1}~Is the HPC community ready (or already late?) to react to the new low-precision formats driven by deep learning?
\MQ{2}~Will HPC navigate itself into a high-precision niche?
\MQ{3}~When if ever, will the industry drop fp64 support?
}

\Resonses{\noindent\begin{minipage}{\linewidth}
    \setcounter{figure}{10}
    \renewcommand{\figurename}{Myth}
    \begin{minipage}[tbp]{0.422\textwidth}
        \centering
        \includegraphics[width=\linewidth]{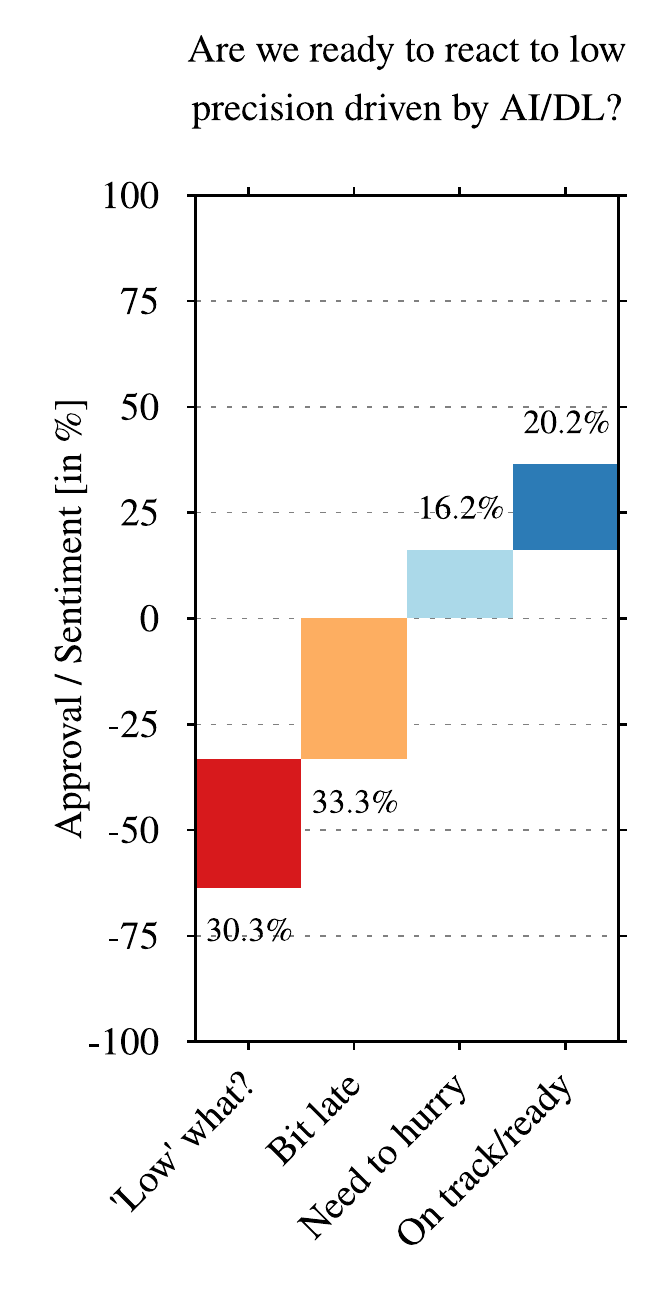}
        \captionof{figure}{Feedback for~\href{https://x.com/thoefler/status/1634089361066033152}{\MQ{1}}}
        \label{fig:myth11}
    \end{minipage}
    \hfill
    \setcounter{figure}{10}
    \begin{minipage}[tbp]{0.528\textwidth}
        \centering
        \includegraphics[width=\linewidth]{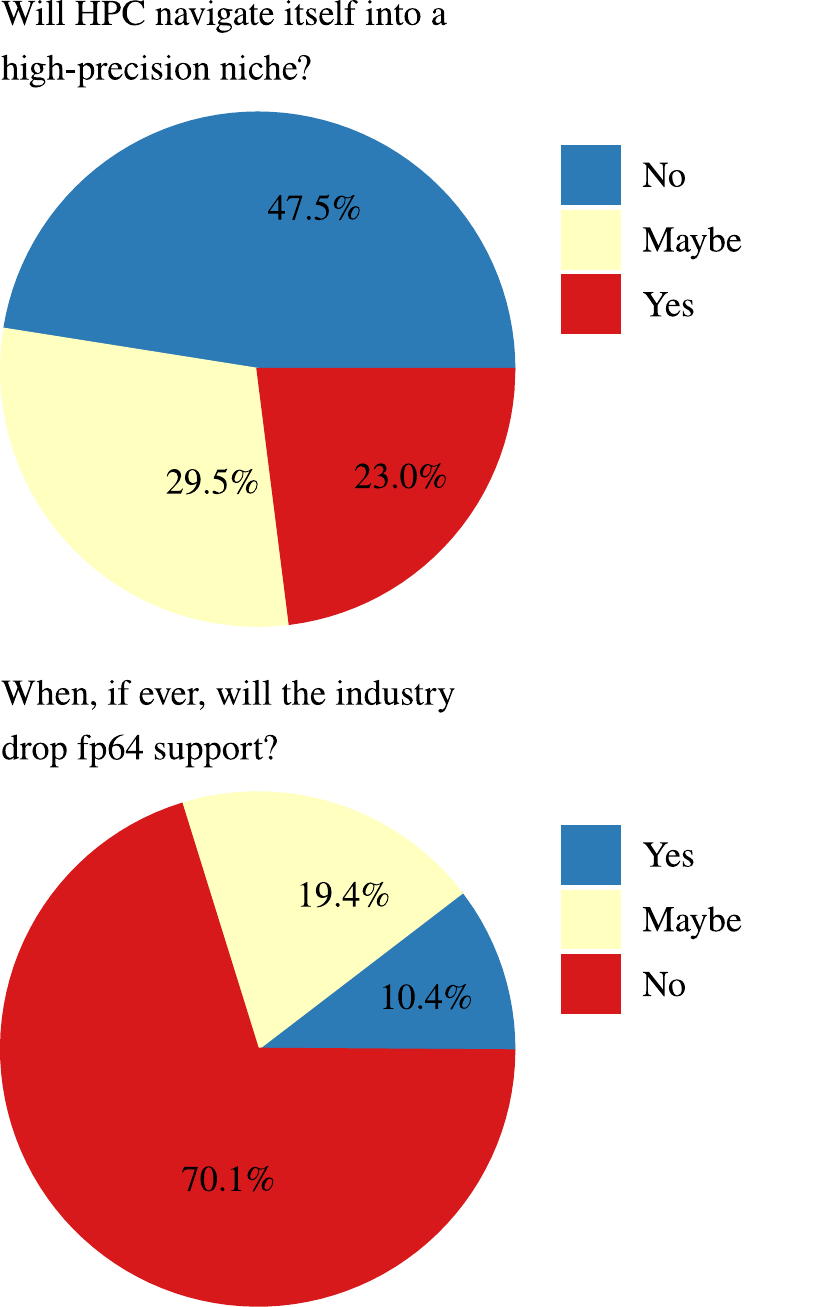}
        \captionof{figure}{Feedback for~\href{https://x.com/thoefler/status/1634089586212196360}{\MQ{2}} \&~\href{https://x.com/thoefler/status/1634089737441951744}{\MQ{3}}}
        \label{fig:myth11b}
    \end{minipage}
\end{minipage}}
\section*{Myth~\nextmyth{sec:clouds}: All HPC Will Be Subsumed by the Clouds!}%\label{sec:clouds}

The rapidly advancing AI and new precision options have reignited the cloud discussion. Whether clouds will subsume supercomputing has been ongoing for more than a decade, since the late 2000s~\cite{deelman_cost_2008}, but remains inconclusive. Today’s cloud offerings offer a wide spectrum for HPC customers, ranging from low-cost standard virtual machines to specialized top-gear HPC equipment in the cloud. It is not surprising that cloud providers offer exactly the same performance as on-premise supercomputing centers in practice~\cite{de_sensi_noise_2022}. After all, they simply buy the same hardware! Thus, this discussion is more of a fiscal argument with an interesting economy-of-scale twist.

There are actually bidirectional aspects to the cloud-vs-supercomputer argument. One is the so-called ``cloudification of supercomputers'', and the latter is ``supercomputification of clouds'', but they often get mixed up leading to confusion in the discussions. We must look at both aspects, and it is in fact the latter where such subsumption may happen or not.

The former, ``cloudification of supercomputers'', is an unmistakable trend, in that various software stack features and APIs are added so that supercomputers effectively become high-end compute resources in the same manner as commercial clouds. Indeed, many major supercomputers are already facilitating cloud features, so that they are effectively clouds themselves, and interoperable with commercial clouds. However, this assumes that there is already a supercomputing resource facilitated for themselves, and does not directly affect the subsumption argument.

The latter, or ``supercomputification of clouds'', is where subsumption may happen, in that clouds nowadays can support features as well as performances of dedicated supercomputers directly, such that they are directly amenable as their replacement. Certainly, there are now multiple cloud services that facilitate virtual compute clusters in the cloud. However, although Intersect 360 reports that the HPC-in-the-cloud compound annual growth rate (CAGR) has been dramatic, over 80\% in 2021~\cite{intersect360_research_worldwide_2022}, it also reports the overall high growth in the HPC market, especially in the high end, and also projects that the growth in the cloud HPC market will flatten over time to be consistent with the overall HPC industry growth. Continued investments by all major global regions in exascale machines and beyond, coupled with companies facilitating their own top-ranked machines, will likely continue to fuel the on-premise infrastructure growth.

In fact, for enterprise IT infrastructures, there has always been a swing between on-premise and public clouds, largely driven by economics. While standing up comprehensive internal IT has become less attractive with multitudes of cloud services readily available in the cloud, so the capital expenditure (CAPEX) for clouds would be cheaper, especially for small enterprises and startups, for large enterprises there is a tendency to move back to on-premise infrastructures, as the operating expenditure (OPEX) of clouds could be expensive. The same could be the case of HPC increasingly as the whole field would pose continuous uprisings in economic viability for industry and societal benefits, thus being driven by economic metrics.

However, the variant of the subsumption scenario is that, although on-premise supercomputers continue to exist, processors and other hardware developments will be largely driven by enterprise HPC needs, currently dominated by AI / deep learning workloads. The R\&D expenditures of hyperscalers in IT now outclass the government investments, and increasingly the hyperscalers are investing in high-end computing. If the commercial cloud hyperscalers can work out the scale of economy in their own hardware manufacturing to the extent that, it could build and operate large-scale HPC infrastructures cheaper than on-premise supercomputers of any size, then the swing could totally happen towards full subsumption— although somewhat unlikely, this could compromise the ability to cover some traditional HPC workloads that do not meet main industrial needs, such as the requirement for dense 64-bit linear algebra capabilities.

\Questions{%
\MQ{1}~When will more than half of the HPC cycles be spent in the cloud?
\MQ{2}~Will on-prem systems be a niche or remain with a significant fraction of HPC cycles spent?
\MQ{3}~What could be a defining development to decide between cloud and on-prem HPC?
}

\Resonses{\noindent\begin{minipage}{\linewidth}
    \setcounter{figure}{11}
    \renewcommand{\figurename}{Myth}
    \begin{minipage}[tbp]{0.422\textwidth}
        \centering
        \includegraphics[width=\linewidth]{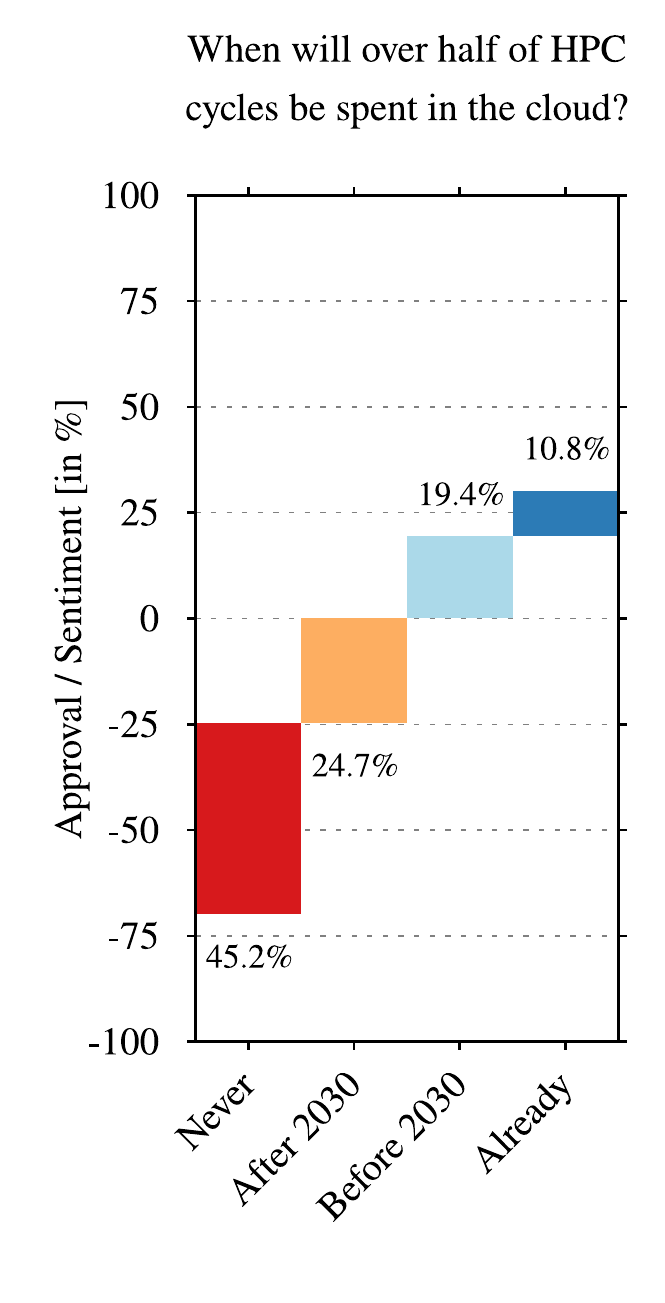}
        \captionof{figure}{Feedback for~\href{https://x.com/thoefler/status/1635267428291846145}{\MQ{1}}}
        \label{fig:myth12}
    \end{minipage}
    \hfill
    \setcounter{figure}{11}
    \begin{minipage}[tbp]{0.560\textwidth}
        \centering
        \includegraphics[width=\linewidth]{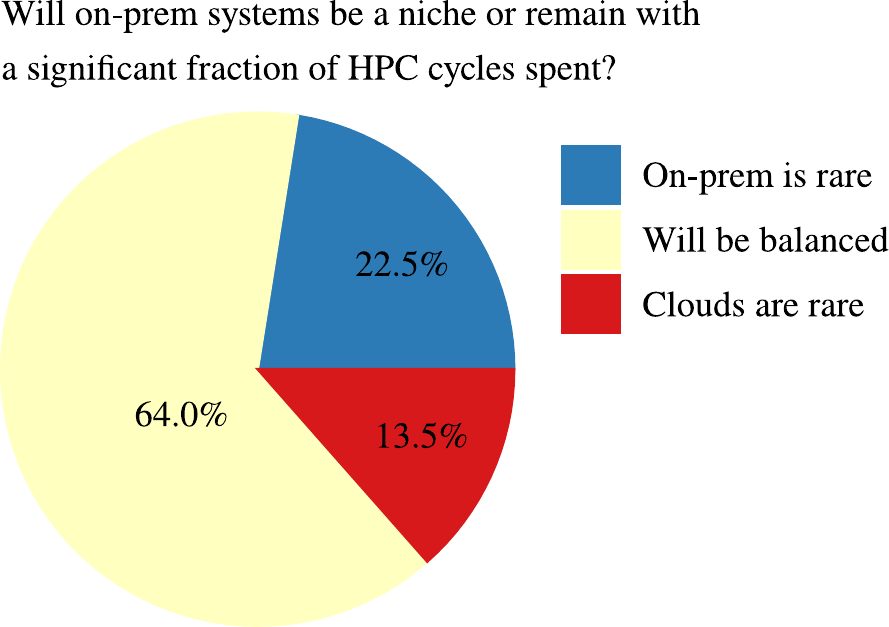}\vspace{10px}
        \includegraphics[width=.6\linewidth]{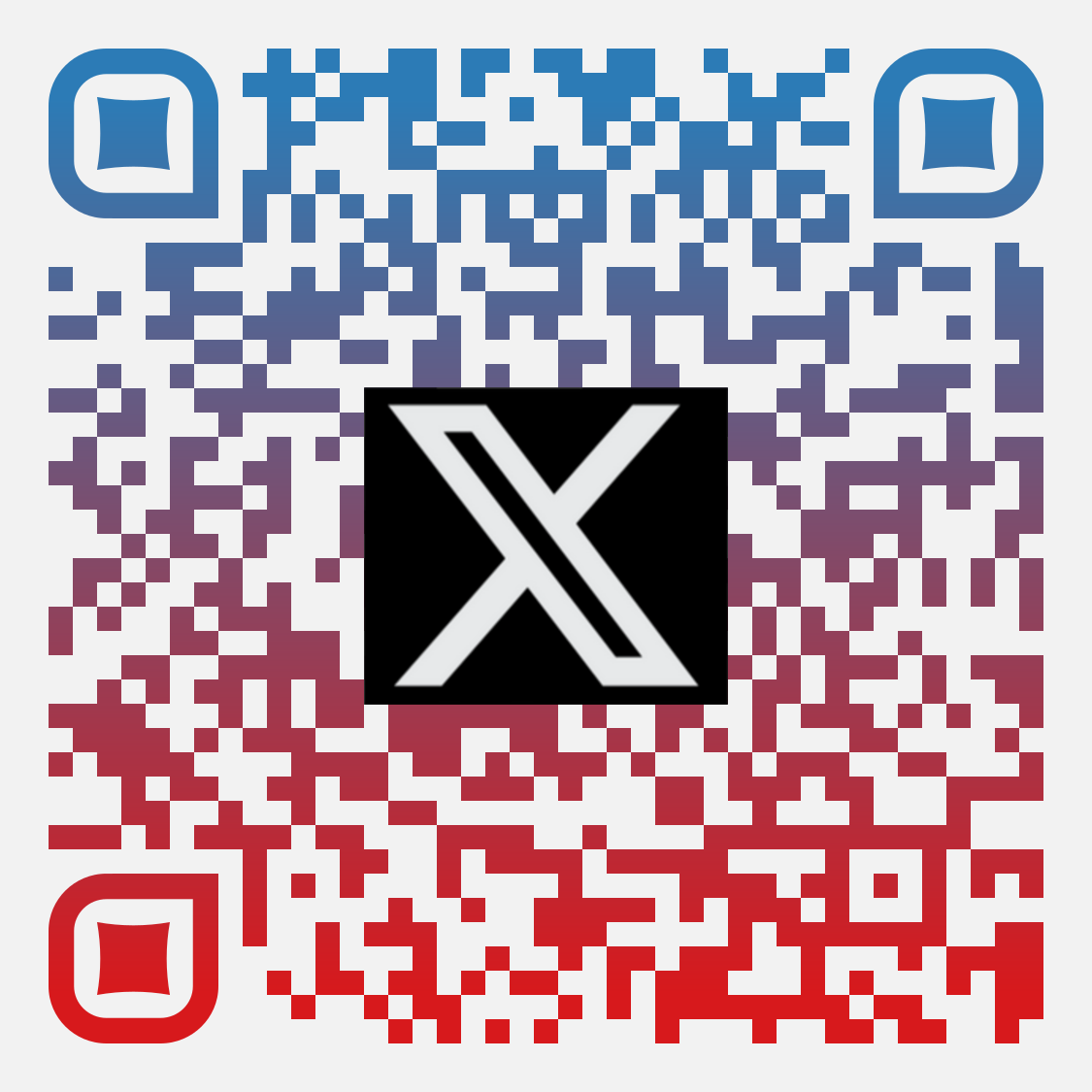}
        \captionof{figure}{Feedback for~\href{https://x.com/thoefler/status/1635267834778644480}{\MQ{2}} and continue discussing~\href{https://x.com/thoefler/status/1635267964273586177}{\MQ{3}} on X}
        \label{fig:myth12b}
    \end{minipage}
\end{minipage}}

\begin{comment}
Given such, here are some of the questions:

Currently, HPC in the cloud mostly covers the low end of the system scaling capabilities, mainly for CAPX reduction of smaller clients. For large clients, it does not make sense due to higher OPEX of clouds. Will there be a scenario in the future that, even for large scale centers with top 100 machines would facilitate their machines virtually in the cloud?

If such happens, the centers could still allocate virtualized supercomputer hardware in the cloud while they preserve the ability to service the client users. Or, the services can also become commercialized for mainstream applications by the ASPs e.g., ReScale\cite{rescale}, to the extent that traditional supercomputer centers will be entirely subsumed. Will that pose problems for the users, or such privatization will not be a hindrance, much the way that most public transportations has been privatized in some way.

If not just the infrastructure, but the hardware that would constitute the infrastructure would be dominated by cloud vendors, will there be significant porations of the current workloads that will be heavily compromised, or will such workloads can largely be accommodate as weill, with improvements in the algorithms / methodologies? For example, if most FP64 compute units are deprecated, will mixed precision arithmetic be sufficient to cover the relative lack of such high precision FP hardware?
\end{comment}

\section*{Conclusions}\label{sec:concl}
\begin{figure}[tbp]
    \setcounter{figure}{3}
    \centering
    \includegraphics[width=0.96\linewidth]{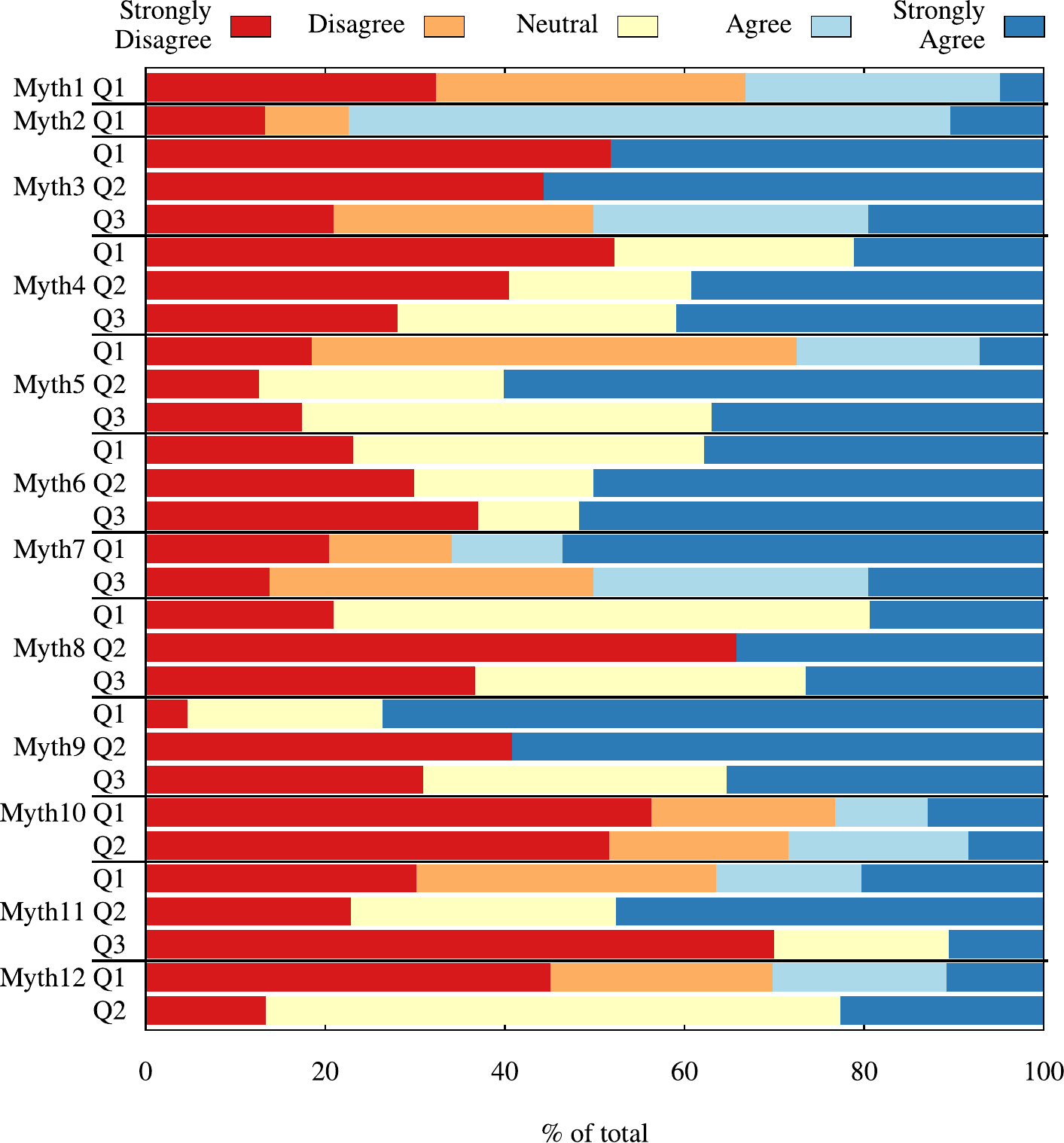}
    \caption{Sentiment analysis of our HPC community conducted via Twitter polls reveals that many do not believe in these myths (in red; strongly disagree on the spectrum) or maintain a pessi- mistic bias toward the posed questions, while a slightly smaller number of pollees are hopeful that these myths become reality.}
    \label{fig:allmyths}
\end{figure}

Many myths shape the discussions in the HPC community today---in this work, we debate some of those myths and hope to stir up arguments. While we present them in an exaggerated and humorous way, many of those myths form the core of thinking in our community. Some may be more divisive than others, but it seems that many are hard to answer definitively. Perhaps some points will be settled in the future, whereas others will not. Yet, their sheer importance mandates serious treatment to help guide future directions for academic research but also industry and government investment.

%To better grasp the divisiveness in our community and to strengthen our anecdotal evidence about these myths, we will conduct large-scale polls within the data center and HPC community over the coming months. Our closing questions from each section will serve as the basis for the polls, and we will summarize the results and update this report~\citep{matsuoka_myths_2023} in due time.

To better grasp the divisiveness in our community and to strengthen our anecdotal evidence about these myths, we conducted large-scale polls within the data center and HPC community in recent months. Our closing questions from each section served as the basis for the polls, and we visualize the individual results thereafter. Across our 33 polls we received 2.557 votes, or an  average of 77 votes per poll (min=36; max=195). For better visualization, we categorize the polls and answers on a sentiment scale; from disagreeing with the myth (typically colored in red) to believing in the myth (blue color). The ones vehemently objecting to the myths have only a small edge with a total of 794 responses across all sentiment polls, while the true believers accumulated 682 responses. If Myth~\ref{sec:fortran} would have been less one-sided, then the number of believers and disbelievers might as well have been near identical. Maybe Fortran will outlive us all despite the many graveside speeches, including the most recent one by~\cite{shipman_evaluation_2023}? 

In Figure~\ref{fig:allmyths}, we summarize the sentiment analysis across all 12 myths and sub-questions. The most notable outliers are likely the disbelieve that Fortran will vanish any time soon, and the believe in reconfigurable hardware (Myth~\ref{sec:fpga}). The latter is also somewhat controversial within the three polled questions---while the majority does not believe in FPGAs for HPC, a similarly majority believes that we have the time \& money to try (and should do so?).

With our community indeed being fractured, only time will tell which of these 12 myths will become true legends!

\section*{Acknowledgements}\label{sec:ackno}

This work was supported by MEXT as ``Feasibility studies for the next-generation computing infrastructure'', NEDO under Grant Number JPNP16007, JST PRESTO Grant Number JPMJPR20MA, JST-CREST under Grant Number JPMJCR19F5, and JSPS KAKENHI Grant Number JP22H03600.

\bibliographystyle{SageH}
\bibliography{main.bib}

\newpage\mbox{}
\newpage
\section*{Authors}

\begin{wrapfigure}{l}{25mm}\
\includegraphics[width=1in,height=1.25in,clip,keepaspectratio]{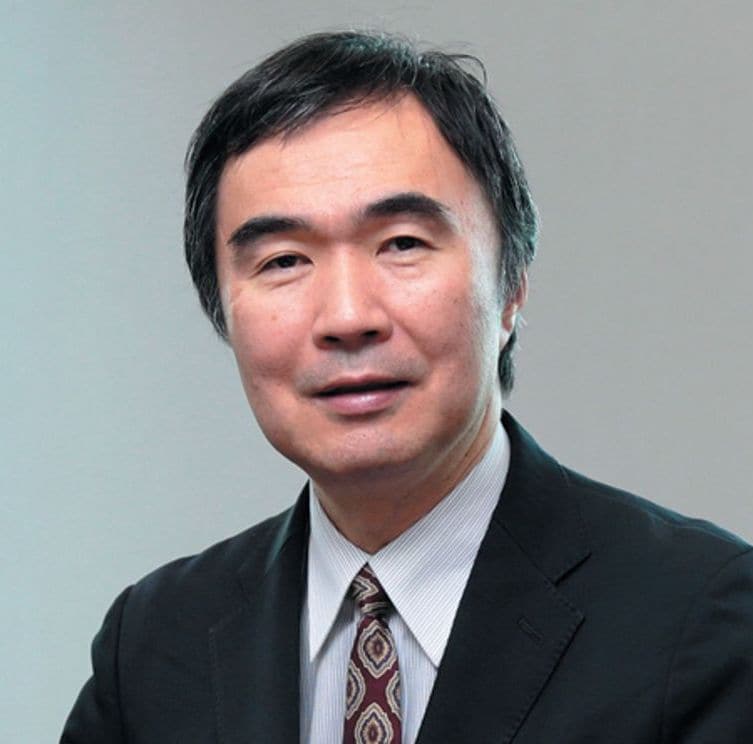}
\end{wrapfigure}
\textbf{Satoshi Matsuoka} from April 2018 has been the director of the RIKEN Center for Computational Science (R-CCS), the top-tier national HPC center for Japan, developing and hosting Japan’s flagship `Fugaku' supercomputer. He was the leader of the TSUBAME series of supercomputers that had also received many international acclaims, at the Tokyo Institute of Technology. His commendations include the Fellow positions in societies/conferences ACM, ISC, and the JSSST; the ACM Gordon Bell Prize in 2011 \& 2021, the IEEE Sidney Fernbach Award in 2014, and the IEEE CS Seymour Cray Award in 2022, all being one of the highest awards in the field of HPC; the Technical Papers Chair and the Program Chair for ACM/IEEE Supercomputing 2009 and 2013 (SC09 and SC13) respectively as well as many other conference chairs, and the ACM Gordon Bell Prize selection committee chair in 2018. He received the Medal of Honor with Purple ribbon by the Japanese government in 2022.

\begin{wrapfigure}{l}{25mm}\
\includegraphics[width=1in,height=1.25in,clip,keepaspectratio]{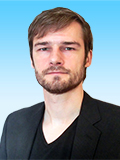}
\end{wrapfigure}\noindent
\textbf{Jens Domke} is the Team Leader of the Supercomputing Performance Research Team at the RIKEN Center for Computational Science (R-CCS), Japan. Jens started his career in HPC in 2008, when he and a team of five students from TU Dresden and Indiana University won the Student Cluster Competition at SC08. Since that time, he has published numerous peer-reviewed journal and conference articles. Jens contributed the DFSSSP and Nue routing algorithms to the subnet manager of InfiniBand, and he built the first large-scale HyperX prototype at the Tokyo Institute of Technology. His research interests include system co-design, performance evaluation, extrapolation, and modelling, interconnect networks, and optimizing parallel applications and architectures. 

\begin{wrapfigure}{l}{25mm}\
\includegraphics[width=1in,height=1.25in,clip,keepaspectratio]{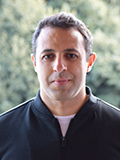}
\end{wrapfigure}\noindent
\textbf{Mohamed Wahib} is the team leader of the ``High Performance Artificial Intelligence Systems Research Team'' at the RIKEN Center for Computational Science (R-CCS), Kobe, Japan. Prior to that he worked as is a senior scientist at AIST/TokyoTech Open Innovation Laboratory, Tokyo, Japan. He received his Ph.D. in Computer Science in 2012 from Hokkaido University, Japan. His research interests revolve around the central topic of high-performance programming systems, in the context of HPC and AI. He is actively working on several projects, including high-level frameworks for programming traditional scientific applications, as well as high-performance AI.

\newpage
\begin{wrapfigure}{l}{25mm}\
\includegraphics[width=1in,height=1.25in,clip,keepaspectratio]{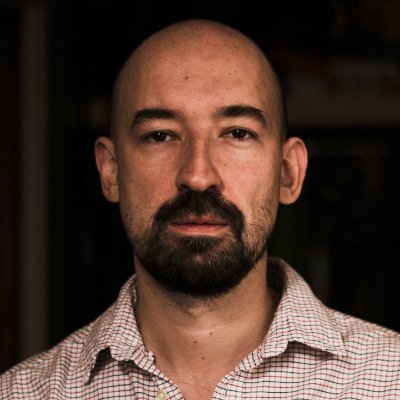}
\end{wrapfigure}\noindent
\textbf{Aleksandr Drozd} is a Research Scientist at the RIKEN Center for Computational Science, High Performance Artificial Intelligence Systems Research Team. His interests lie at the intersection of high performance computing and artificial intelligence, particularly in areas such as learning and evaluating text representations. Aleksandr holds a Ph.D. degree from the Department of Mathematical and Computing Sciences at the Tokyo Institute of Technology, Japan.

\begin{wrapfigure}{l}{25mm}\
\includegraphics[width=1in,height=1.25in,clip,keepaspectratio]{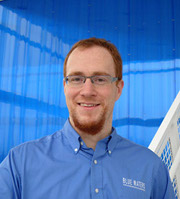}
\end{wrapfigure}\noindent
\textbf{Torsten Hoefler} is a Professor of Computer Science at ETH Zurich, a member of Academia Europaea, and a Fellow of the ACM and IEEE. His research interests revolve around the central topic of ``Performance-centric System Design'' and include scalable networks, parallel programming techniques, and performance modeling. Torsten won best paper awards at the ACM/IEEE Supercomputing Conference SC10, SC13, SC14, SC19, SC22, EuroMPI'13, HPDC'15, HPDC'16, IPDPS'15, and other conferences. He published numerous peer-reviewed scientific conference and journal articles and authored chapters of the MPI-2.2 and MPI-3.0 standards. He received the IEEE CS Sidney Fernbach Award, the ACM Gordon Bell Prize, the Latsis prize of ETH Zurich, as well as both ERC starting and consolidator grants. Additional information about Torsten can be found on his homepage at \url{htor.inf.ethz.ch}.

%\section{Copyright statement}
%Please  be  aware that the use of  this \LaTeXe\ class file is
%governed by the following conditions.
%
%\subsection{Copyright}
%Copyright \copyright\ \volumeyear\ SAGE Publications Ltd,
%1 Oliver's Yard, 55 City Road, London, EC1Y~1SP, UK. All
%rights reserved.
%
%\subsection{Rules of use}
%This class file is made available for use by authors who wish to
%prepare an article for publication in a \textit{SAGE Publications} journal.
%The user may not exploit any
%part of the class file commercially.
%
%This class file is provided on an \textit{as is}  basis, without
%warranties of any kind, either express or implied, including but
%not limited to warranties of title, or implied  warranties of
%merchantablility or fitness for a particular purpose. There will
%be no duty on the author[s] of the software or SAGE Publications Ltd
%to correct any errors or defects in the software. Any
%statutory  rights you may have remain unaffected by your
%acceptance of these rules of use.
%
%\begin{acks}
%This class file was developed by Sunrise Setting Ltd,
%Brixham, Devon, UK.\\
%Website: \url{http://www.sunrise-setting.co.uk}
%\end{acks}
%
%\begin{thebibliography}{99}
%\bibitem[Kopka and Daly(2003)]{R1}
%Kopka~H and Daly~PW (2003) \textit{A Guide to \LaTeX}, 4th~edn.
%Addison-Wesley.
%
%\bibitem[Lamport(1994)]{R2}
%Lamport~L (1994) \textit{\LaTeX: a Document Preparation System},
%2nd~edn. Addison-Wesley.
%
%\bibitem[Mittelbach and Goossens(2004)]{R3}
%Mittelbach~F and Goossens~M (2004) \textit{The \LaTeX\ Companion},
%2nd~edn. Addison-Wesley.
%
%\end{thebibliography}

\end{document}